\documentclass[11pt]{article}
\usepackage{epsf,amsmath,amsfonts,epsfig,theorem,version}
\usepackage{graphics}
\usepackage{float}
\usepackage{authblk}
\usepackage{hyperref,listings}
\RequirePackage[OT1]{fontenc}
\usepackage{subcaption}
\usepackage{caption}
\DeclareCaptionFont{figureCaptionSize}{\fontsize{9}{7.3}\selectfont}
\usepackage{cite}

\usepackage{nameref,hyperref}

\usepackage[right]{lineno}

\usepackage{microtype}
\DisableLigatures[f]{encoding = *, family = * }

\usepackage{tikz}

\usepackage{changepage}

\usepackage[aboveskip=1pt,labelfont=bf,labelsep=period,singlelinecheck=off]{caption}

\makeatletter
\renewcommand{\@biblabel}[1]{\quad#1.}
\makeatother

\usepackage{lastpage,fancyhdr,graphicx}
\usepackage{epstopdf}
\usepackage{color}

\definecolor{Gray}{gray}{.25}

\usepackage{graphicx}

\usepackage{sidecap}

\usepackage{wrapfig}
\usepackage[pscoord]{eso-pic}
\usepackage[fulladjust]{marginnote}
\reversemarginpar

\definecolor{dkgreen}{rgb}{0,0.6,0}
\definecolor{gray}{rgb}{0.5,0.5,0.5}
\definecolor{mauve}{rgb}{0.58,0,0.82}

\newcommand{\be}{\begin{equation}}
\newcommand{\ee}{\end{equation}}
\newcommand{\bes}{\begin{equation*}}
\newcommand{\ees}{\end{equation*}}

\begin{document}
\title
{\Large
\textbf{New numerical approaches for modeling thermochemical convection in a compositionally stratified fluid}
}

\author[1]{
Elbridge Gerry Puckett\footnote{
egpuckett@ucdavis.edu
}}
\author[2]{
Donald L Turcotte}
\author[2,3]{
Louise H Kellogg}
\author[1]{\\
Ying He}
\author[1]{
Jonathan M.~Robey}
\author[3]{
Harsha Lokavarapu}

\affil[1]{\small Department of Mathematics, U.\ C. Davis, Davis, CA 95616, USA.}
\affil[2]{\small  Department of Earth and Planetary Science, U.\ C. Davis, Davis, CA 95616, USA
}
\affil[3]{ \small Computational Infrastructure for Geodynamics, U.\ C. Davis, Davis, CA 95616, USA}

\date{}
\maketitle

\begin{abstract}
Seismic imaging of the mantle has revealed large and small scale heterogeneities in the lower mantle; 
specifically structures known as large low shear velocity provinces (LLSVP) below Africa and the South Pacific. 
Most interpretations propose that the heterogeneities are compositional in nature, differing in 
composition from the overlying mantle, an interpretation that would be consistent with chemical geodynamic models.
The LLSVP's are thought to be very old, meaning they have persisted throughout much of Earth's 
history.
Numerical modeling of persistent compositional interfaces presents challenges, even to
state-of-the-art numerical methodology. 
For example, some numerical algorithms for advecting the compositional interface cannot
maintain a sharp compositional boundary as the fluid migrates and distorts with time dependent 
fingering due to the numerical diffusion that has been added in order to maintain the upper and 
lower bounds on the composition variable and the stability of the advection method.
In this work we present two new algorithms for maintaining a sharper computational boundary
than the advection methods that are currently openly available to the computational mantle 
convection community; namely, a Discontinuous Galerkin method with a Bound 
Preserving limiter (DGBP) and a Volume-of-Fluid (VOF) interface tracking algorithm. 
We compare these two new methods with two approaches commonly used for modeling the 
advection of two distinct, thermally driven, compositional fields in mantle convection 
problems; namely, an approach based on a high-order accurate finite element method (FEM)
advection algorithm that employs an artificial viscosity technique known as `Entropy 
Viscosity' (FEM-EV) to maintain the upper and lower bounds on the composition variable as 
well as the stability of the advection algorithm and the advection of particles that carry a 
scalar quantity representing the location of each compositional field.
All four of these algorithms are implemented in the open source FEM code ASPECT, which we use 
to compute the velocity, pressure, and temperature fields associated with the underlying flow 
field.
We compare the performance of these four algorithms by computing the solution to an initially 
compositionally stratified fluid that is subject to a thermal gradient at a Rayleigh number 
of $\mathrm{Ra} = 10^5$ with a buoyancy ratio of $\mathrm{B} = 1.0$. 
For $\mathrm{B} = 1.0$, a value for which the initial stratification of the compositional 
fields persists indefinitely, our computations demonstrate that the entropy viscosity-based 
method has  far too much numerical 
diffusion to yield meaningful results. 
On the other hand, the DGBP method yields good results, although small amounts of each 
compositional field are numerically entrained within the other compositional field.
The particle method yields yet better results than this, but some particles 
representing the denser fluid are entrained in the upper, less dense fluid and are advected 
to the top of the computational domain and, similarly, particles representing  the less dense 
fluid are entrained in the lower, denser fluid and are advected to the bottom of the 
computational domain.
The VOF method maintains a sharp interface between the two compositions on a subgrid scale 
throughout the computation.
We also compute the same problem with $\mathrm{Ra} = 10^5$ for a range of buoyancy ratios
$\mathrm{B} \, = \, 0.0, \, 0.1, \, 0.2, \, \ldots , \, 1.0$ using the DGBP method in order to 
demonstrate the utility of this method when the stratified layer overturns and kinematic mixing begins.
In this regime the VOF algorithm is not suitable for modeling the interface between 
the two compositional fields, since this would require that the grid size $h \to 0$ 
as the curvature $\sigma$ of the interface between the two compositional fields goes to 
infinity, $\sigma \to \infty$. 

\end{abstract}

\vskip 12pt


\thispagestyle{fancy}

\section{Introduction}

A major unresolved question in geodynamics is whether
mantle convection consists of a single layer or two
separate layers. Seismic studies have shown deep
penetration of subducted lithosphere which favors whole
mantle convection. However isotopic studies provide
strong evidence for a well mixed upper mantle but a
lower mantle reservoir containing primordial material and
unmixed subducted material. The near uniform isotopic ratios of most mid-ocean ridge basalts 
indicate that this volcanism is sampling a near homogeneous upper-mantle reservoir.
However, the extreme isotopic heterogeneities of ocean island basalts indicate that there is a 
deep source.
This source includes primordial materials as well as a variety of subducted materials that have 
not been mixed into the mantle.
The transport of mantle from the deep source is attributed to mantle plumes \cite{kellogg1999compositional,turcotte2014geodynamics}.

Recent studies utilizing seismic imaging have revealed large regions with anomalous seismic 
properties in the lower mantle. There are two dome like regions beneath Africa and the Pacific 
with low shear-wave velocities extending some 1000 km above the core-mantle boundary with 
horizontal dimensions of several thousand kilometers. These are known as large low shear-velocity 
provinces (LLSVPs).
Most interpretations propose that the heterogeneities are compositional in nature, differing from 
the overlying mantle, an interpretation that would be consistent with chemical geodynamic models. 
The LLSVP’s are thought to be very old, meaning they have persisted throughout much of Earth's 
history \cite{Burke200849,Cottaar2012213,french2015broad}.

Mantle convection with compositional differences, together with a thermal contribution, is 
known as \text{thermal-chemical convection}. 
\cite{tackley2012dynamics} has given a comprehensive review of the influence of compositional 
buoyancy on mantle convection both in terms of evidence for buoyancy structures in the mantle and 
models of the influence of compositional buoyancy on thermal-chemical convection. 
Thermal-chemical convection with applications to the mantle requires distinct regions with 
different compositions and densities.
Molecular diffusion that can mix regions with different compositions can occur only on very small 
scales in the mantle on geological time scales, centimeters to meters.
Thus, mixing is predominantly kinematic with sharp compositional 
interfaces~\cite{kellogg1992mixing}.

Numerical modeling of persistent compositional interfaces presents challenges, even to 
state-of-the-art numerical methodology.
Currently, in the computational mantle convection community many numerical algorithms for 
modeling distinct compositional regions do not maintain sharp boundaries between distinct 
compositions as they migrate and distort with time dependent fingering. 
In particular, some of these algorithms exhibit compositional diffusion, either as an 
unwanted \textit{numerical artifact} of the algorithm or \textit{artificial diffusion}, 
which is an explicit design feature of the algorithm, or both.
Artificial diffusion, which is typically referred to as artificial viscosity, is a quantity that is 
added to the (numerical) advection equation in order to maintain the upper and lower bounds on the 
value of the compositional variable and also, for Finite Element Methods (FEM), to maintain the 
stability of the advection algorithm~\cite{vonneumann1950method}, \cite{guermond2011entropy}.
This numerical technique is described in more detail in
Section~\ref{Subsubsection:Finite Element Advection Algorithm with Entropy Viscosity}.
Note that we prefer to use the term artificial diffusion instead of artificial viscosity.
Numerical algorithms that allow the compositional boundary to diffuse, whether as a 
\textit{numerical artifact} or as an intentional part of the advection method are inherently 
in conflict with the fact that the true sharp compositional boundary must persist indefinitely.

A number of studies of the role of compositional discontinuities on mantle convection have 
been carried out.
Numerical studies include \cite{montague2000numerical}, \cite{mcnamara2004thermochemical}, 
\cite{tan2005metastable}, and \cite{galsa2015effective}.
\cite{davaille1999two} has carried out an extensive laboratory studies.

In this paper we exam four alternative algorithms for numerically modeling this compositional 
discontinuity.
All of these algorithms are implemented as part of the open source mantle convection software
\href{https://geodynamics.org/cig/software/aspect/}{\textsc{ASPECT}}, which is described in 
detail in~\cite{MK-TH-WB:2012}.
Each algorithm is designed to model the motion of two distinct compositional regions.
These four algorithms are: 
1) the compositional advection method with so-called `Entropy Viscosity' (FEM-EV), which was 
the first advection method to be implemented in ASPECT,
2) a new Discontinuous Galerkin Bound Preserving (DGBP) compositional advection method, 
3) Volume-of-Fluid (VOF) interface tracking, and 
4) Particles.
These algorithms will be described in detail in
Sections~\ref{Subsubsection:Finite Element Advection Algorithm with Entropy Viscosity}--\ref{Subsubsection:Particles}
and their relative performance will be examined and discussed in
Section~\ref{Section:Numerical Results} below.

\vskip 48pt 

\begin{table}[h!]
    \centering
    \caption{A list of symbols used in this paper.}
    \label{Table:List of Symbols}
    \scalebox{0.85}{
    \begin{tabular}{lll || lll}
        \hline
        Symbol     & Quantity                  & Unit             & Symbol        & Quantity                  & Unit \\
        \hline
        $\bf{u}$   &Velocity                   & $m / s$               & $\rho$        &Density                    & $kg\cdot m^{-3}$\\
        $p$        &Dynamic pressure           & Pa               &  $\Delta\rho$ &Density difference   & $kg\cdot m^{-3}$    \\
        $T_0$      &Temperature at the top     & K                & $D$           & Compositional diffusivity  &  $m^2/s$              \\
        $T_1$      &Temperature at the bottom  & K                & $\alpha$      & Thermal expansion coefficient &  1/K               \\
        $T$        &Temperature                & K                & $d$           & Vertical thickness of fluid layer & m \\
        $\Delta T$ &Temperature difference     & K                & $Pr$          &Prandtl number    & $\frac{\mu}{\rho\kappa}$    \\
        $C$        &Composition                & -                & $Le$          &Lewis number     & $\frac{\kappa}{D}$    \\
        $\mu$      &Viscosity                  & Pa $\cdot$ s     & $Ra$          &Rayleigh number  & $\frac{\rho_0{\bf{g}}\alpha\Delta Td^3}{\mu\kappa}$\\
        $\kappa$   &Thermal diffusivity        & $m^2/s$          & $B$           &Buoyancy ratio & $\frac{\Delta\rho}{\rho_0\alpha\Delta T}$         \\ 
        $\rho_0$   &Reference density          & $kg\cdot m^{-3}$ &  & & \\    
        \hline
    \end{tabular}
    }
\end{table}

\section{The Model Problem}
\label{Section:The Model Problem}

\subsection{The Dimensional Form of the Equations}

In order to study alternative numerical algorithms for modeling persistent compositional 
interfaces we will consider a problem that emphasizes the effect of a compositional density 
difference on thermal convection.
We consider a two-dimensional flow in a horizontal fluid layer with a thickness $d$.
In dimensional terms our problem domain $\Omega$ has width $3d$ and depth $d$.
At a given reference temperature $T_0$ the region $0 \le y < d / 2$ has a compositional 
density of $\rho_0$ and the region $d / 2 < y \le d$ has a compositional density of 
$\rho_0 + \Delta \rho$ where $\Delta \rho \, \ll \, \rho_0$.

We also introduce a composition variable $C(x, y, t)$ defined  by
\begin{equation}
  \label{Def:Composition Variable C}
    C = \frac{\rho - \rho_0}{\Delta \rho} \, .       
\end{equation}
The composition $C$ is the concentration of the dense fluid as a function of space and time.
The initial condition for $C$ is 
\begin{equation}
  \label{Eq:Initial Conditions for C}
    C( x, y, t=0) \, = \,
      \left\{
        \begin{array}{ll}
          0 & \text{for} \quad 0     \leq y \leq d / 2 \, , \\
          1 & \text{for} \quad d / 2   <  y \leq d     \, .
       \end{array}
     \right.
\end{equation}

The upper boundary, at $y = 0$, has temperature $T_0$ and the lower boundary at $y = d$ has
temperature $T_1$.
The fluid is assumed to have a constant viscosity $\mu$ which is large. 
The Prandtl number 
\begin{equation}
    Pr \, = \, \frac{\mu}{\rho_0 \kappa} \, \gg \, 1 \, ,
\end{equation}
where $\kappa$ is the thermal diffusivity so that inertial effects can be neglected.
The fluids in the high density and low density layers are immiscible; i.e., they cannot mix by 
diffusion. 
The Lewis number 
\begin{equation}
    Le \, = \, \frac{\kappa}{D} \, \gg \, 1 \, ,
\end{equation}
where $D$ is the diffusion coefficient for the compositional variable $C$
(Table~\ref{Table:List of Symbols}). 
Thus, the discontinuous boundary between the high density and low density fluids is preserved 
indefinitely.  

The problem we have posed requires the solution of the standard equations for thermal 
convection with the addition of an equation for the compositional field $C$ that 'tracks' 
the compositional density.
We make the assumption that the Boussinesq approximation holds; namely, that density 
differences associated with convection $\rho_0 \alpha(T_1 - T_0)$ and $\Delta\rho$ are
small compared with the reference density $\rho_0$.
\begin{equation}
  \label{Eq:Dimensional Boussinesq approximation}
    \rho( x, y, t) \, = \, \rho_0 \, ( 1 - \alpha (T - T_0)) \, + \, \Delta \rho \, C \, .
\end{equation}  

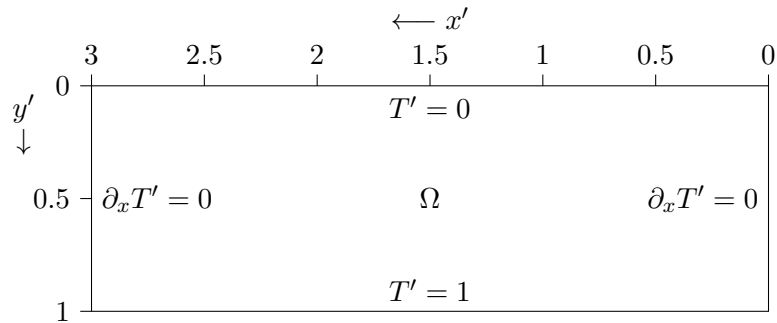
\begin{figure}[h!]
  \centering
    \begin{tikzpicture}[scale=3]
    \draw (0,0) rectangle (-3,-1);
    \foreach \x in {0, 0.5, 1, 1.5, 2, 2.5, 3}
    \draw (-\x, 0) -- (-\x, 0.05) node[anchor=south] {$\x$};
    \foreach \y in {0, 0.5, 1}
    \draw (-3, -\y) -- (-3.05, -\y) node[anchor=east] {$\y$};
    \node at (-1.5, -0.5) {$\Omega$};
    \node at (-1.5, 0.3) {$\longleftarrow x^{\prime}$};
    \node at (-3.3, -0.1) {$y^{\prime}$};
    \node at (-3.3, -0.25) {$\downarrow $};
    \node[anchor=west]  at (-3, -0.5) {$\partial_xT^{\prime} = 0$};
    \node[anchor=east]  at (0, -0.5)  {$\partial_xT^{\prime} = 0$};
    \node[anchor=north] at (-1.5, 0)  {$T^{\prime} = 0$};
    \node[anchor=south] at (-1.5, -1) {$T^{\prime} = 1$};
    \end{tikzpicture}
    \caption{The geometry of the (nondimensional) computational domain $\Omega$ shown with 
        the temperature boundary conditions on the four side walls. 
        The velocity boundary conditions on the side walls are
        $\mathbf{u} \cdot \mathbf{n}=0$ (no flow) and
        $\partial \bf{u} / \partial \boldsymbol{\tau} = 0$ (free slip) where $\mathbf{n}$ 
        and $\boldsymbol{\tau}$ are the unit normal and tangential vectors to the boundary 
        respectively.
            }
  \label{Fig:The Nondimensional Domain}
\end{figure}

The governing equations have been discussed in detail by~\cite{schubert2001mantle} 
(Chapter~6). 
Conservation of mass requires
\begin{equation}
  \label{Eq:Dimensional Conservation of Mass}
    \frac{\partial u}{\partial x} + \frac{\partial v}{\partial y} = 0
\end{equation}  
where $x$ and $y$ denote the horizontal and vertical spacial coordinates, oriented as shown 
in Figure~\ref{Fig:The Nondimensional Domain}, and $u$ and $v$ denote the horizontal and 
vertical velocity components, respectively.
We use the Stokes equations
\begin{equation}
  \begin{aligned}
    \label{Eq:Dimensional Stokes Equation for u}
       0 &= \frac{-\partial P}{\partial x} 
            + \mu(\frac{\partial^2 u}{\partial x^2}
            + \frac{\partial^2 u}{\partial y^2}) \, ,
  \end{aligned}
\end{equation}

\begin{equation}
\begin{aligned}
    \label{Eq:Dimensional Stokes Equation for v}
      0 &= \frac{-\partial P}{\partial y}
           + \mu(\frac{\partial^2 v}{\partial x^2}
           + \frac{\partial^2 v}{\partial y^2}) 
           - \rho_0 \alpha (T - T_0) g + \Delta\rho C g \, ,
\end{aligned}
\end{equation}
where $P$ is the dynamic pressure, $\alpha$ is the coefficient of thermal expansion, and 
$g$ is the gravitational acceleration in the positive (downward) $y$ direction as shown in 
Figure~\ref{Fig:The Nondimensional Domain}.
\noindent
Conservation of energy requires 
\begin{equation}
  \label{Eq:Dimensional Temperature Equation}
    \frac{\partial T}{\partial t} + u\frac{\partial T}{\partial x}
      + v\frac{\partial T}{\partial y} 
      = \kappa(\frac{\partial^2 T}{\partial x^2} + \frac{\partial^2 T}{\partial y^2})\,.
\end{equation}

With no diffusion, i.e., $D = 0$, the composition variable $C$ satisfies the
\textit{advection equation} 
\begin{equation}
  \label{Eq:Dimensional Composition Equation}
    \frac{\partial C}{\partial t}
      + u\frac{\partial C}{\partial x} + v\frac{\partial C}{\partial y}
      = 0\,.
\end{equation}

\subsection{The Nondimensional Form of the Equations}
We introduce the non-dimensional variables 
\begin{equation}
  \begin{aligned}
    \label{Eq:Nondimensionalized Variables}
    x' &= \frac{x}{d},               & y' &= \frac{y}{d},            & t' &= \frac{\kappa}{d^2} \, t, \\
    u' &= \frac{d}{\kappa} \, u,     & v' &= \frac{d}{\kappa} \, v,      & \rho' &= \frac{\rho}{\rho_0},   \\
    T' &= \frac{T - T_0}{T_1 - T_0}, & P^\prime &= \frac{d^{\, 2} P}{\mu \kappa},
  \end{aligned}
\end{equation}
and the two nondimensional parameters, the Rayleigh number $Ra$ and the buoyancy ratio $B$
\begin{equation}
  \label{Def:Rayleigh Number}
    Ra = \frac{\rho_0 \, g \, \alpha(T_1 - T_0)d^3}{\mu \, \kappa},
\end{equation}

\begin{equation}
  \label{Def:Buoyancy Ratio}
    B \, = \, \frac{\Delta\rho}{\rho_0 \alpha(T_1 - T_0)}.
\end{equation}
Substitution of equations~\eqref{Eq:Nondimensionalized Variables}--\eqref{Def:Buoyancy Ratio} 
into
equations~\eqref{Eq:Dimensional Conservation of Mass}--\eqref{Eq:Dimensional Composition Equation}
gives
\begin{align}
   \label{Eq:Dimensionless Continuity Equation}
    \frac{\partial u'}{\partial x'} + \frac{\partial v'}{\partial y'}
    &= 0                                                                              \, , \\
  \label{Eq:Dimensionless Stokes Equations For u}
    0 &= \frac{-\partial P'}{\partial x'}
       + \frac{\partial^2 u'}{\partial x^{'2}}
       + \frac{\partial^2 u'}{\partial y^{'2}}                                        \, , \\
  \label{Eq:Dimensionless Stokes Equations For v}
    0 &= \frac{-\partial P'}{\partial y'}
       + \frac{\partial^2 v'}{\partial x^{'2}}
       + \frac{\partial^2 v'}{\partial y^{'2}}
       - \mathrm{Ra} T' + \mathrm{Ra} \mathrm{B} C                                    \, , \\
  \label{Eq:Dimensionless Temperature Advection Diffusion Equation}
    \frac{\partial T'}{\partial t'}
      + u' \frac{\partial T'}{\partial x'} 
      + v' \frac{\partial T'}{\partial y'}
     &= \frac{\partial^2 T'}{\partial x^{'2}} + \frac{\partial^2 T'}{\partial y^{'2}} \, , \\
  \label{Eq:Dimensionless Composition Equation}
    \frac{\partial C}{\partial t'} + u'\frac{\partial C}{\partial x'}
                                   + v'\frac{\partial C}{\partial y'}
     &= 0 \, .
\end{align}

This is the superposition of a Rayleigh-Taylor problem and a Rayleigh-Bernard problem. In 
the isothermal limit ($T_0 = T_1$) it is the classic Rayleigh-Taylor problem 
(\cite{turcotte2014geodynamics}, pp. 285-286).
If $C$ is positive, a light fluid is above the heavy fluid and in a downward gravity field 
the fluid layer is stable. 
If $\Delta\rho$ is negative, a heavy fluid lies over a light fluid and the layer is 
unstable. 
Flows will transfer the heavy fluid to the lower half and the light fluid to the upper 
half. 
The density layer will overturn. 
If $\Delta \rho = 0$ this is the classic Rayleigh-Bernard problem for thermal convection. 
The governing parameter is the Rayleigh number. 
If $0 < Ra < Ra_{c}$, the critical Rayleigh number, no flow will occur. 
If $Ra_c < Ra < Ra_t$ steady cellular flow will occur. 
If $Ra > Ra_t$ the flow becomes unsteady and thermal turbulence develops.  

If $Ra > Ra_c$ and $B$ is small, the boundary between the density differences will not 
block the flow driven by thermal convection.
Kinematic mixing will occur and the composition will homogenize so that the density is 
constant and $C = \frac{1}{2}$. Whole layer convection will occur.
If B is large, the density difference boundary will block the flow driven by thermal 
convection.
The compositional boundary will be displaced vertically but will remain intact.  Layered 
convection will occur with the compositional boundary, the boundary between the convecting 
layers.  
The Rayleigh number defined in equation~\eqref{Def:Rayleigh Number} is based on the domain 
thickness $d$.
This is the case for which we will show numerical computations.
\section{The Numerical Methodology}
\label{Section:The Numerical Methodology}

In the following discussion of the numerical methodology, we will only consider the 
dimensionless 
equations~\eqref{Eq:Dimensionless Continuity Equation}-\eqref{Eq:Dimensionless Composition Equation}
and drop the primes associated with the dimensionless variables.  
The vector form of the dimensionless equations on the 2D rectangular domain  
$\Omega=[0,3]\times[0,1]$ shown in Figure~\ref{Fig:The Nondimensional Domain} are given by
\begin{align}
  \label{Eq:Nondimensional Vector Form of the Stokes Equations}
   -\nabla^2 {\bf{u} }+\nabla  P   &= (-Ra T + Ra BC) \, {\bf{g}} \\
     \label{Eq:Nondimensional Vector Form of the Continuity Equation}
    \nabla\cdot{\bf {u}} &= 0, \\
 \label{Eq:Nondimensional Vector Form of the Temperature Equation}
    \frac{\partial T}{\partial t}+{\bf u}\cdot\nabla T &= \nabla^2 T,\\
     \label{Eq:Nondimensional Vector Form of the Composition Equation}
    \frac{\partial C}{\partial t}+{\bf u}\cdot\nabla C &=0,
\end{align}
where  ${\bf u}=(u,v)$, and ${\bf g}= (0, 1)$ is the unit vector pointing downward.
We also assume no-flow and free-slip velocity boundary conditions on all boundaries,
\begin{align}
\label{Eq:No-FLow BCs}
\mathbf{u} \cdot \mathbf{n}                         &= 0 
\qquad \text{(no-flow)}   \, , \\
\label{Eq:Free-Slip BCs}
\frac{\partial \bf{u}}{ \partial \boldsymbol{\tau}} &= 0
\qquad \text{(free slip)} \, ,
\end{align}
where $\mathbf{n}$ and $\boldsymbol{\tau}$ are the unit normal and tangential vectors to the 
boundary respectively.
We impose Dirichlet boundary conditions for the temperature on the top and bottom of the 
computational domain and Neumann boundary conditions (no heat flux)  on the sides of the computational domain,
\begin{eqnarray}
  \label{Eq:Temperature BCs y eq 0} 
    T ( x, 0, t) = 0             \, ,   \\
  \label{Eq:Temperature BCs y eq 1}
     T ( x, 1, t) = 1             \, ,  \\
  \label{Eq:Temperature BCs x eq 0}
    \partial_{x} T ( 0, y, t) = 0  \, , \\
  \label{Eq:Temperature BCs x eq 3}
  \partial_{x} T ( 0, y, t) = 0 \, .
\end{eqnarray}
The geometry of the computational domain together with the boundary conditions 
on the temperature are shown in Figure~\ref{Fig:The Nondimensional Domain}.

\subsection{Decoupling of the Nonlinear System}
\label{Cubsection:Nonlinear Decoupling}
The incompressible Stokes equations can be considered as a constraint on the temperature and 
composition at any given time leading to a highly nonlinear system of equations.
To solve this nonlinear system, we apply the Implicit Pressure Explicit Saturation (IMPES) 
approach, which was originally developed for computing approximations to solutions of 
equations for modeling problems in porous media flow 
\cite{sheldon1959one,huber1999multiphase}, to decouple the incompressible Stokes
equations~\eqref{Eq:Nondimensional Vector Form of the Stokes Equations}
--\eqref{Eq:Nondimensional Vector Form of the Continuity Equation}
from the temperature and compositional 
equations~\eqref{Eq:Nondimensional Vector Form of the Temperature Equation}--\eqref{Eq:Nondimensional Vector Form of the Composition Equation}. 
This leads to three discrete systems of linear equations, the Stokes equations, the temperature 
equation and the composition equation, thereby allowing them to be solved easily and efficiently. 

\subsection{Discretization of the Stokes Equations}
Let $t_k$ denote the discretized time at the k{\it th} time step with a time step size
$\Delta t_k =t_k - t_{k-1}$, $k = 0, \, 1, \, \ldots$ 
Given the temperature $T^k$ and composition $C^k$ at time $t = t_k$, we first solve for our approximation to the Stokes
equations~\eqref{Eq:Nondimensional Vector Form of the Stokes Equations}
--\eqref{Eq:Nondimensional Vector Form of the Continuity Equation} to obtain the 
velocity  $\mathbf{u}^k = (u^k , \, v^k)$ and pressure $P^k$
\begin{align}
  \label{Eq:Stokes Discretization a}
   -\nabla^2 {\bf{u} }^k +\nabla  P^k   &= (-Ra T^k +RaBC^k){\bf{g}}, \\
   \label{Eq:Stokes Discretization b}
                   \nabla\cdot{\bf {u}}^k &= 0. 
\end{align}
For the incompressible Stokes
equations~\eqref{Eq:Stokes Discretization a}-\eqref{Eq:Stokes Discretization b}, we use the 
standard mixed FEM method with a Taylor-Hood element \cite{Donea2005} for the spatial 
approximation of the incompressible Stokes
equations~\eqref{Eq:Nondimensional Vector Form of the Stokes Equations}--\eqref{Eq:Nondimensional Vector Form of the Continuity Equation}.
If we assume the finite element space is $(X(\Omega))^n\times Y(\Omega)$, where $n$ denotes the 
spacial dimension then the problem is to find 
$({\bf u}^k,P^k)\in(X(\Omega))^n\times Y(\Omega)\cap (H^1_0(\Omega))^n \times L_2(\Omega)$
such that 
\begin{align}
  \label{Eq:Weak Formulation of Stokes Equations}
    \int\limits_\Omega \nabla{\bf{u}}^k\cdot\nabla {\bf{\Phi}}dxdy
     -\int\limits_\Omega  P^k\nabla\cdot{\bf {\Phi}}dxdy &= \int
    \limits_\Omega(-Ra T^k +RaBC^k){\bf g}\cdot{\bf\Phi}dxdy,       \\
  \label{Eq:Weak Formulation of the Continuity Equation}
    \int\limits_\Omega \nabla\cdot{\bf{u}}^k \, q \, dx\, dy &= 0
\end{align}
for any
$({\bf \Phi},q)\in (X(\Omega))^n \times Y(\Omega) \cap (H^1_0(\Omega))^n \times L_2(\Omega)$.
If we introduce the inner product of two vector functions $\mathbf{u}$ and $\mathbf{v}$ on a given 
domain $\Omega$
\begin{align}
    (\mathbf{u},\mathbf{v})_{\Omega}=\int\limits_\Omega {\bf{u}}\cdot {\bf{v}}dxdy,
\end{align}
and the inner product of two scalar functions $u$ and $v$ on $\Omega$
\begin{align}
    (u,v)_{\Omega} = \int\limits_\Omega \, u \,v \, dx \, dy
\end{align}
then we can rewrite the 
equations~\eqref{Eq:Weak Formulation of Stokes Equations}--\eqref{Eq:Weak Formulation of the Continuity Equation}
as  
\begin{align}
  \label{Eq:wkstokes_v2}
    ( \nabla{\bf{u}}^k,\nabla {\bf{\Phi}})_{\Omega} -( P^k,\nabla\cdot{\bf {\Phi}})_{\Omega}
   &=(Ra(B \ C^k - T^k){\bf g},{\bf\Phi})_{\Omega},\\
  \label{Eq:wkstokes_div_v2}
  ( \nabla\cdot{\bf{u}}^k,q)_{\Omega} &= 0 \, .
\end{align}

\subsection{Discretization of the Temperature Equation}
In all of the computations presented here we use the algorithm currently implemented in ASPECT to 
approximate the spatial and temporal terms in the temperature
equation~\eqref{Eq:Nondimensional Vector Form of the Temperature Equation}.
This algorithm includes the entropy viscosity stabilization technique described 
in~\cite{JLG-RP-BP:2011} and~\cite{MK-TH-WB:2012}.
The weak form of this spatial discretization is
\begin{align}
  \label{Eq:Weak Form of the Spatial Discretization of the Temperature Equation}
    (\frac{\partial T}{\partial t},\psi)_{\Omega}+({\bf u}\cdot\nabla T,\psi)_{\Omega}
      &=-(\nabla T,\nabla\psi)_{\Omega}-(\nu_h(T)\nabla T,\nabla\psi)_{\Omega}  
       + (\frac{\partial T}{\partial {\bf n}},\psi)_{\Gamma_D}
\end{align}
where  $\nu_h(T)$ is the entropy viscosity function as defined in~\cite{MK-TH-WB:2012}, 
except here we do not use a second-order extrapolation to treat the advection term
$({\bf u}\cdot \nabla T,\psi)$ and the entropy viscosity term  
$(\nu_h(T)\nabla T,\nabla\psi)_{\Omega}$ explicitly.
The entropy-viscosity function $\nu_h^k(T)$ is a non-negative constant within each cell that only 
adds artificial diffusion in cells for which the local P\'{e}clet number 
$\mathrm{Pe} = \mathrm{Ra} \ \mathrm{Pr}$ is large and the solution is not smooth.
We use the fully implicit adaptive Backward Differentiation Formula of order 2 (BDF2) 
\cite{wanner1991, MK-TH-WB:2012} to discretize the temperature equation in time.
Thus, the full discretization of the temperature equation is
\begin{align}
  \label{Eq:Weak Form of the Discretization of the Temperature Equation}
    &(\frac{1}{\Delta t_{k+1}}\left (\frac{2\Delta t_{k+1}+\Delta t_{k}}{\Delta t_{k+1}+\Delta t_{k}}T^{k+1}-\frac{\Delta t_{k+1}+\Delta t_{k}}{\Delta t_{k}}T^k+\frac{\Delta t_{k+1}^2}{\Delta t_{k}(\Delta t_{k+1}+\Delta t_{k})}T^{k-1}\right ),\psi)_{\Omega}
       \nonumber \\
   =&-({\bf u}^{k}\cdot\nabla T^{k+1},\psi)_{\Omega}-(\nabla T^{k+1},\nabla\psi)_{\Omega}-(\nu_h^k(T)\nabla T^{k+1},\nabla\psi)_{\Omega}   
      +(\frac{\partial T^{k+1}}{\partial {\bf n}},\psi)_{\Gamma_D}\,.
\end{align}

\subsection{Discretization of the Composition Equation}
\label{Subsection:Discretization of the Composition Equation}

We use one of the four algorithms described below to discretize the composition 
equation~\eqref{Eq:Nondimensional Vector Form of the Composition Equation}.

\subsubsection{The Finite Element Advection Algorithm with Entropy Viscosity}
\label{Subsubsection:Finite Element Advection Algorithm with Entropy Viscosity}

This is the first advection algorithm that was implemented in ASPECT.
It is based on the same spatial discretization as shown
in equation~\eqref{Eq:Weak Form of the Spatial Discretization of the Temperature Equation}.
However, the entropy-viscosity stabilization term on the right-hand side in 
\begin{align}
  \label{Eq:C_FEM}
    (\frac{\partial C}{\partial t},\psi)_{\Omega}
      + ({\bf u}\cdot\nabla C,\psi)_{\Omega} 
      \, = \, -(\nu_h^k(C) \nabla C, \nabla \psi)_{\Omega}
\end{align}
is computed separately for the composition field; i.e, it does \textit{not} have the same
value in each cell as does the entropy-viscosity $\nu_h^k(T)$ for the temperature field.
In equation~\eqref{Eq:C_FEM} the entropy function $\nu_h^k(C)$ has the same purpose as 
$\nu_h^k(T)$ and is defined by 
\begin{align}
  \label{Def:Entropy Viscosity}
    \nu^e_h
     = \frac{\alpha \, h^2 \, R( \,r_e(C_h))}{ \|E(C_h)-\bar E\|_{\infty,\Omega}} \, .
\end{align}
See \cite{JLG-RP-BP:2011},\cite{MK-TH-WB:2012} for details. 
We also use the adaptive BDF2 algorithm for the time discretization, leading to the following 
FEM Entropy Viscosity (FEM-EV) discretization
of equation~\eqref{Eq:Nondimensional Vector Form of the Composition Equation},
\begin{align}
  \label{Eq:The Entire FEM-EV Discretization}
      ( \frac{1}{\Delta t_{k+1}} &\left (\frac{2\Delta t_{k+1}+\Delta t_{k}}{\Delta 
          t_{k+1}+\Delta t_{k}}C^{k+1}-\frac{\Delta t_{k+1}+\Delta t_{k}}{\Delta t_{k}}C^k+\frac{\Delta t_{k+1}^2}{\Delta t_{k}(\Delta t_{k+1}+\Delta t_{k})}C^{k-1} \right),\psi )_{\Omega} \nonumber\\
     =& - ({\bf u}^{k}\cdot\nabla C^{k+1},\psi)_{\Omega}-(\nu_h^k(C)\nabla C^{k+1},\nabla\psi)_{\Omega} \, .
\end{align}

\subsubsection{The Discontinuous Galerkin Bound Preserving Advection Algorithm}
\label{Subsubsection:The Discontinuous Galerkin Bound Preserving Advection Algorithm}

In this algorithm we use adaptive BDF2 to discretize the advection
equation~\eqref{Eq:Nondimensional Vector Form of the Composition Equation} for the 
composition in time as shown in equation~\eqref{Eq:The Entire FEM-EV Discretization}.
However we use a Discontinuous Galerkin method with a Bound Preserving limiter (DGBP) for the 
discretization of the spatial terms
in equation~\eqref{Eq:Nondimensional Vector Form of the Composition Equation}.
The DG method differs from the classic continuous Galerkin FEM, since it allows for discontinuities 
between elements~\cite{WHR-TRH:1973, CWS:2016}.
If we denote the discretized computational domain by $\Omega=\cup_{e=1}^E\Omega^e$, where 
$\Omega^e$ denotes non-overlapping body-conforming quadrilateral elements, and let 
$V_N(\Omega)$ denote the DG element space, then fully discretized problem is as follows.
Find $C^{k+1}\in V_N(\Omega)$ such that
\begin{align}
  \label{Eq:FULL_C_DG}
      &(\frac{1}{\Delta t_{k+1}}\left (\frac{2\Delta t_{k+1}+\Delta t_{k}}{\Delta t_{k+1}+\Delta t_{k}}C^{k+1}-\frac{\Delta t_{k+1}+\Delta t_{k}}{\Delta t_{k}}C^k+\frac{\Delta t_{k+1}^2}{\Delta t_{k}(\Delta t_{k+1}+\Delta t_{k})}C^{k-1}\right ),\psi)_{\Omega_e}\nonumber\\
   =&-({\bf u}^{k}\cdot\nabla C^{k+1}, w)_{\Omega_e}+({\bf u}^{k}\cdot {\bf n}    
   C^{k+1},w)_{\partial\Omega_e}
    -({\bf u}^{k}\cdot {\bf n} C^{k+1,*},w)_{\partial\Omega_e}
\end{align}
for any $w \in V_N$.
An upwind flux is used to determine $C^{k+1,*}$,
\begin{align}
  C^{k+1,*} =
    \left\{
      \begin{matrix}
        C^{k+1,-},&\text{ if }{\bf u^{j}  \cdot n} > 0\,, \\
        C^{k+1,+},&\text{ if } {\bf u^{j} \cdot n} < 0\,, \\
      \end{matrix}
    \right.
\end{align}
where $C^{k+1,-}$ is the local/interior solution on $\Omega_e$, and $C^{k+1,+}$ is the 
neighbor / exterior solution of $\Omega_e$~\cite{cockburn1998local,hesthaven2008nodal}. 
Although, equation \eqref{Eq:FULL_C_DG} appears to be only defined locally on each element 
$\Omega_e$, it also depends on the adjacent solutions through the flux term 
$C^{k+1,*}$, which is defined at each element interface using two side values.
Generally, the flux term $C^{k+1,*}$ is the most difficult part to determine when one wants to 
design a DG method, as it is the essential feature of the algorithm that ensures the
stability of the method.

After obtaining the DG solution, we apply a Bound Preserving (BP) limiter, which was 
initially developed in~\cite{XXZ-CWS:2010}, and further developed for approximating solutions 
of the equations for modeling convection in the Earth's mantle in~\cite{YH-EGP-MIB:2016}. 
See the latter reference for a detailed explanation of the DGBP algorithm that we have used in 
this work.

\subsubsection{The Volume-of-Fluid Interface Tracking Algorithm}
\label{Subsubsection:The Volume-of-Fluid Interface Tracking Algorithm}
The advection methods described in
Sections~\ref{Subsubsection:Finite Element Advection Algorithm with Entropy Viscosity} and
in~\ref{Subsubsection:The Discontinuous Galerkin Bound Preserving Advection Algorithm} 
above are often referred to as \textit{interface capturing} methods, since the interface 
between the two compositions is not explicitly tracked.
In contrast, the Volume-of-Fluid (VOF) method is an \textit{interface tracking} method 
in which, at each time step, the interface is explicitly reconstructed
\textit{in every cell that contains a portion of the interface} and the interface is 
advanced in time using the explicit knowledge of the interface location and topology at 
the current time step. 
Note that this means the interface is being resolved on a sub-grid scale.

In a typical VOF algorithm one discretizes the equation
\begin{equation}
  \label{Eq:Conservation Equation for the Volume Fraction f}
    \frac{\partial f}{\partial t} \,  + \,\nabla\cdot \mathbf{F}\left( f \right) \, = \, 0,
\end{equation}
where $\mathbf{u} = ( u,v)$ is the velocity field and $f$ denotes the
\textit{volume fraction} of one of the compositional fields, say the field with density 
$\rho_0 \, + \, \Delta \rho$, which we will refer to as
composition~01\footnote[1]{Usually $f$ is referred to the volume fraction of `fluid 01' or 
    `material 01', etc. Hence the name \textit{Volume}-of-Fluid'.}, and 
\begin{equation}
  \label{Def:The Volume Fraction Flux}
   \mathbf{F}\left( f \right) 
     \, = \, \left( F \left( f \right) , G (f) \right)
     \, = \, \left( u \, f , \, v \, f \, \right)  
\end{equation}
is the flux associated with composition~01.
For example, let $\Omega_e$ denote an arbitrary finite element cell in our domain
and let $f^k_e$ denote the (discretized) volume fraction in $\Omega_e$ at time $t_k$.
Then the \textit{volume} of composition~01, $V_e$, in $\Omega_e$ at time $t_k$ is 
\begin{equation}
  \label{Eq:The Volume of Composition 01 in Omega_e}
    V_e \, = \, \int\limits_{\Omega_e} \, f^k_e \, dx \, dy \, .
\end{equation}
Note that for an incompressible velocity field $\mathbf{u} = ( u,v)$ we have
$\nabla \cdot \mathbf{u} = 0$ and hence, the volume of `parcels' of composition~01 are 
constant as they evolve in time.
Consequently the volume of composition~01 is conserved over time.
In particular, equation~\eqref{Eq:Conservation Equation for the Volume Fraction f} is a 
conservation equation for $f$.
From a more mathematical point of view, $f(x,y)$ is the characteristic function associated
with composition~01,
\begin{equation}
     f(x,y) \, = \, 
     \begin{cases}
         f(x,y) \, &= \, 1 \quad \text{if $(x,y)$ is occupied by composition~01}  \, , \\
         f(x,y) \, &= \, 0 \quad \text{if $(x,y)$ is not occupied by composition~01} \, .
  \end{cases}
\end{equation}

In its  simplest form our implementation of the VOF algorithm in ASPECT proceeds as 
follows.\footnote[2]{For simplicity of exposition we will assume the grid consists of 
square grid cells $\Omega_e$ aligned parallel to the $x$ and $y$ axes.}
Given the values $f^k_e$ at time $t_k$ and the velocity field at $t_k$ and $t_{k+1}$ 
we do the following:
\begin{enumerate}
    
    \item (The Interface Reconstruction Step) Given the $f^k_e$ on all grid cells 
           $\Omega_e$ \textit{reconstruct} the interface
          in $\Omega_e$.

    \item (The Advection Step) Given the reconstructed interface in $\Omega_e$ and the 
          velocity normal to the edges of $\Omega_e$ at time $t_{k+{\frac{1}{2}}}$, 
          determine the flux $\mathbf{F}^{k+\frac{1}{2}}_e \left( f \right)$ at the half
          time step $t_{k+{\frac{1}{2}}}$ of composition~01 across each of the edges of $\Omega_e$.
          
     \item (The Conservative Update) Given the flux
           $\mathbf{F}^{{k+\frac{1}{2}}}_e \left( f \right)$ at the half time 
           time step $t_{k+\frac{1}{2}}$ update the values of the volume fraction 
           in each of the cells $\Omega_e$,
           \begin{equation}
             \label{Eq:Conservative VOF Update}
               f^{k+1}_e \,
                 = \,  f^k_e \, 
                   + \, \frac{\Delta t_k}{h} \,
                   \left[F^{k+\frac{1}{2}}_L - F^{k+\frac{1}{2}}_R
                     + \,[G^{k+\frac{1}{2}}_B - G^{k+\frac{1}{2}}_T\,] \right]
           \end{equation}
           where $F^{k+\frac{1}{2}}_L$, $F^{k+\frac{1}{2}}_R$, $G^{k+\frac{1}{2}}_T$
           $G^{k+\frac{1}{2}}_B$ denote the fluxes across the left, right, top, and bottom
           edges of $\Omega_e$ as depicted in Figure~\ref{Fig:VOF Conservative Update}.
                  
\end{enumerate} 
The fluxes are evaluated at the half time level $t_{k+\frac{1}{2}}$ in order to obtain 
second-order accuracy

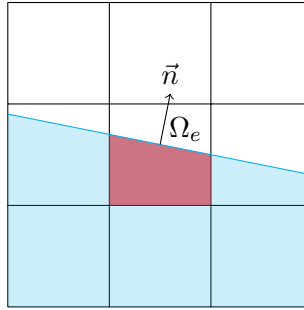
\begin{figure}[tbph!]
  \centering
    \begin{tikzpicture}[scale=1.35]
   \draw (-1,-1) grid (2,2);
   \node at (0.75, 0.75) {$\Omega_e$};
\fill[color=red,opacity=0.6] (0,0) -- (1,0) -- (1,0.5) -- (0,0.7);
\draw[color=cyan] (-1,0.9) -- (2,0.3);
\fill[color=cyan,opacity=0.2] (-1,-1) -- (-1,0.9) -- (2,0.3) -- (2,-1);
\draw[->] (0.5,0.6) -- (0.6,1.1) node[above] {\(\vec{n}\)};
\end{tikzpicture}
    \caption{In our implementation of the VOF algorithm the interface is approximated as a 
            line segment in each cell $\Omega_e$ that contains a volume fraction $f_{e}$ 
            with $0 < f_{e} < 1.$
            The algorithm we use has been proven to be a second-order accurate 
            approximation to a smooth curve provided that the condition 
            in equation~\eqref{Def:VOF EGP Constraint} holds.}
    \label{Fig:The VOF Reconstruction Algorithm}
\end{figure}

In the reconstruction step~(1) most VOF codes reconstruct (approximate) the interface in 
$\Omega_e$ as a line.
In our VOF implementation in ASPECT we use the `Efficient Least Squares VOF Interface 
reconstruction Algorithm' (ELVIRA), which is described in detail in  
\cite{JEP-EGP:2004} and is based on the ideas in \cite{EGP:1991} and \cite{JEP:1992}.
In the ELVIRA interface reconstruction algorithm we use the information in the cells 
immediately adjacent to the cell $\Omega_e$ in which we wish to reconstruct the interface
to determine a linear approximation to the interface as depicted in 
Figure~\ref{Fig:The VOF Reconstruction Algorithm}.
In~\cite{EGP:2010a} and \cite{EGP:2014}it was proved that this algorithm produces a 
second-order accurate approximation to the interface provided that
\begin{equation}
  \label{Def:VOF EGP Constraint}
    h \, \le \, \frac{2}{33 \, \sigma_{max}}
\end{equation}
where  $\sigma_{max}$ denotes the maximum curvature of the interface and $h$ is the grid 
size.
 
\begin{figure}[tbph!]
  \centering
    \begin{tikzpicture}[scale=1.0]
    \draw[<->] (-1.5,0.5) node[left] {$y$} -- (-1.5,-0.5) -- (-0.5,-0.5) node[below] {$x$};
    \draw (0,0) rectangle (4,4);
    \node at (1.5, 3.0) {$\Omega_e$};
    \draw[dashed] (0,0) rectangle (4,-4);
    \fill[color=red,opacity=0.5] (0,0) -- (4,0) -- (4,1) -- (0,3);
    \draw[color=green] (4,4) -- (2,2.6) -- (2,-1.4) -- (4,0);
    \fill[color=red,opacity=0.6] (2,2) -- (2,0) -- (4,0) -- (4,1);
    \fill[color=purple,opacity=0.6] (2,0) -- (2,-1.4) -- (4,0);
    \draw[line width=1pt, ->] (4,2) -- (5.0,2)  node[anchor=west] {$F^{k+\frac{1}{2}}_R$};
    \draw[line width=1pt, ->] (5.5,0.2) -- (6.5,0.85) node[anchor=west] {$\mathbf{u}^{k+\frac{1}{2}}_R$};
    \end{tikzpicture}
    \caption{A \textit{geometric} depiction of the computation of the flux $F^{k+\frac{1}{2}}_R$ 
             of composition~01 across the right-hand edge of a square grid cell $\Omega_e$.
             The volume of composition~01 with its reconstructed interface in $\Omega_e$ 
             is colored red.
             The intersection of this region with the characteristics traced back from the
             right-hand corners of $\Omega_e$ is dark red.
             The volume of this region is a second-order accurate approximation to the 
             \textit{volume} flux of composition~01 across the right-hand  edge of 
             $\Omega_e$.
             `Higher \textit{resolution}', but still second-order accurate, approximations of 
             this flux are possible.
             For example, if  the purple region is included in the computation of the flux
             this `\textit{high resolution}' second-order computation of the flux is known 
             as `Corner Transport Upwind'~\cite{PC:1990, RJL:1996}.
             }
    \label{Fig:VOF Computation of the Flux}
\end{figure}
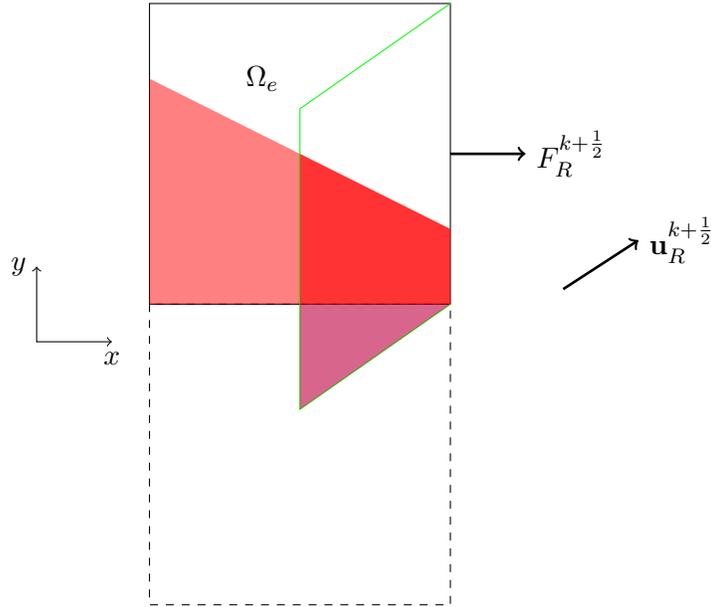
\begin{figure}[tbph!]
  \centering
     \begin{tikzpicture}
  \draw (0,0) rectangle (4,4);
  \node at (2.75, 2.75) {$\Omega_e$};
  \draw[<->] (-1,0) node[anchor=south] {$y$} -- (-1,-1) -- (0, -1) node[anchor=west] {$x$};

  \draw[line width=1pt, ->] (0,2) -- (0.75,2) node[anchor=west] {$F^{k+\frac{1}{2}}_L$};

  \draw[line width=1pt, ->] (4,2) -- (4.75,2) node[anchor=west] {$F^{k+\frac{1}{2}}_R$};

  \draw[line width=1pt, ->] (2,0) -- (2,0.75) node[anchor=south] {$G^{k+\frac{1}{2}}_B$};

  \draw[line width=1pt, ->] (2,4) -- (2,4.75) node[anchor=south] {$G^{k+\frac{1}{2}}_T$};
\end{tikzpicture}
    \caption{A depiction of the fluxes $F^{k+\frac{1}{2}}_L$, $F^{k+\frac{1}{2}}_R$,  $
             G^{k+\frac{1}{2}}_B$, and $G^{k+\frac{1}{2}}_T$, that we use in step (3), the 
             VOF conservative update. 
             This depiction assumes a square grid cell with uniform side of length $h$.
            }
    \label{Fig:VOF Conservative Update}
\end{figure}
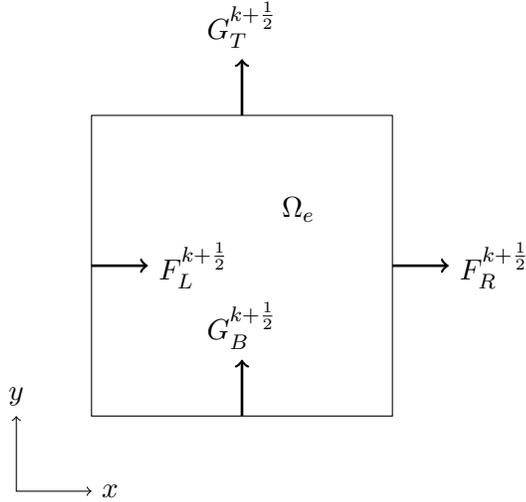
In step (2) the volume of composition 01 that will cross each edge is computed geometrically 
by intersecting it with the region that is formed by tracing each edge back along the 
characteristics as shown for the right-hand edge of $\Omega_e$ in
Figure~\ref{Fig:VOF Computation of the Flux}.
In this figure the boundary of this region is shown in green and the intersection of this region 
with the region containing composition 01 is shown in dark red.

Finally, given the value of the four fluxes $F^{k+\frac{1}{2}}_L$, $F^{k+\frac{1}{2}}_R$, 
$G^{k+\frac{1}{2}}_B$, and $G^{k+\frac{1}{2}}_T$ obtained in step (2) as described above, we use
equation~\eqref{Eq:Conservative VOF Update} to update the value of the volume fraction
$f^{k}_e \, \to \, f^{k+1}_e$.

This is a simplified explanation of the structure of a VOF algorithm.
The details of its implementation in ASPECT are more complicated and will be the subject of a
paper by the fifth and first authors.

\subsubsection{Particles}
\label{Subsubsection:Particles}

Particle methods, sometimes called `Particle-In-Cell' or `Tracer in Cell' methods -- as well 
as other names -- have long been used by researchers to model problems involving convection 
in the Earth's Mantle; e.g.,~ \cite{PJT-SDK:2003}, \cite{mcnamara2004thermochemical}, 
\cite{SJT-PJH-CS-JPL:2014}, and \cite{trim2016interaction}.
The accuracy of high-order accurate versions of these methods have been recently studied, 
both in conjunction with Finite Difference~\cite{TD-DAM-TVG-PJT:2011} and Finite Element 
methods~\cite{thielmann2014discretization}, as well as their efficient 
parallel implementation in ASPECT~\cite{RG-EH-EGP-WB:2016}.
The particle algorithm in ASPECT that we use here is based on a second-order Runge-Kutta time 
discretization and an arithmetic averaging interpolation algorithm.
It is described in detail in~\cite{RG-EH-EGP-WB:2016} and \cite{TH-JD-RG-WG:2017}.

All three of the algorithms discussed above are based on the Eulerian frame of 
reference, or Eulerian coordinates.
Therefore, the flow velocity field $\bf u$ is represented as a vector function of position 
$\bf x$ and time $t$, ${\bf u}={\bf u}({\bf{x}},t)$.
However, since the individual fluid particles are followed as they move through the grid over 
time it is convenient to use a Lagrangian frame of reference in order to have a convenient 
notation for describing the location of the particles in time. 
In particular, we will label each particle with a (time-independent) vector $\bf x_0$, which is
the initial location of the particle at time $t = 0$.
We use the vector function ${\bf X}({\bf x_0},t)$ to denote the location of the particle 
with initial position ${\bf x_0}$ at time $t$.
Thus,  ${\bf X}({\bf x_0},t)$ satisfies the following equation
\begin{align}
  \label{Eq:Particles}
    \frac{\partial {\bf X}({\bf x_0},t)}{\partial t}={\bf u}({\bf X}({\bf x_0},t),t)
\end{align}
where ${\bf u}({\bf x},t)$ is the velocity at the point ${\bf x}$ at time $t$.
Therefore, given the initial positions of the particles ${\bf x_0}$, we can solve 
equation~\eqref{Eq:Particles} to evolve the particle locations in time. 

Now denote the discretization of the particles in time by 
${\bf X}^{k}_{\bf x_0}\approx {\bf X}({\bf x_0},t_{k})$ and
${\bf u}^{k}({\bf X}^{k}_{\bf x_0})\approx {\bf u}({\bf X}^{k}_{\bf x_0},t_{k})$, 
We use a second order Runge Kutta time discretization to approximate
equation~\eqref{Eq:Particles}
\begin{align}
  \label{Eq:C_PIC}
   {\bf X}^{k+1,1/2}_{\bf x_0}&={\bf X}^k_{\bf x_0}+\frac{1}{2}\Delta t_{k+1}{\bf u}^{k}({\bf X}^{k}_{\bf x_0}),\\
    {\bf X}^{k+1}_{\bf x_0}&={\bf X}^{k}_{\bf x_0}+\Delta t_{k+1}{\bf u}\left({\bf X}^{k+1,1/2}_{\bf x_0} ,t_k+\frac{1}{2}\Delta t_{k+1}\right )\,,
\end{align}
where we approximate the velocity at the half time $t_k+\frac{1}{2}$ by averaging the velocities at $t_k$ and $t_{k+1}$
\begin{equation}
   {\bf u}\left({\bf X}^{k+1,1/2}_{\bf x_0} ,t_k+\frac{1}{2}\Delta t_{k+1}\right )
     \approx \frac{1}{2}\left\{{\bf u}^{k+1}({\bf X}^{k+1,1/2}_{\bf x_0})+{\bf u}^{k}({\bf X}^{k+1,1/2}_{\bf x_0})\right\} \, .
\end{equation}

Once we obtain the new particle locations ${\bf X}^{k+1}_{\bf x_0}$  at time $t_{k+1}$, we 
compute the value of the composition `carried' by that particle
$C^{k+1}({\bf X}^{k+1}_{\bf x_0})$ by assigning it the value it had at time $t = 0$; 
i.e.,  $C({\bf x_0})$. 
We use ASPECT to compute the approximate velocity field ${\bf u}({\bf x},t_{k})$ at each time 
step.
In order to project the compositional value carried by each particle onto the quadrature points 
of a given finite element cell $\Omega_e$ at time $t_{k}$ , we use the arithmetic average of the 
values of the particles in $\Omega_e$,
\begin{align}
  C^k_{e} = \frac{\sum _p C_p}{\sum_p 1}
\end{align}
where $p$ is the index of each particle located within $\Omega_e$.
The value of the compositional variable $C$ at each of the quadrature points in $\Omega_e$ 
at time $t_{k}$ is simply the constant $C^k_e$.


\section{Numerical Results}
\label{Section:Numerical Results}
The domain for all of the computational results shown below is a rectangular two-dimensional 
box, which we denote by $\Omega= [0,3]\times[0,1]$ as shown in
Figure~\ref{Fig:The Nondimensional Domain}. 
We use the following initial conditions for the temperature
\begin{align}
  \label{Eq:Dimensionaless Temperature Initial Condition}
    T(x,y;t=0) =
      \left\{
        \begin{array}{ll}
          y-A\sin(\pi y)(1-\cos(2\pi x)) & \text{if}   \quad 0\le y\le \frac{1}{10} \, , \\
          y+A\sin(\pi y)(1-\cos(2\pi x)) & \text{if}   \quad \frac{9}{10}\le y\le 1 \, , \\
          0.5                            & \phantom{if}\quad \text{otherwise}       \; , \\
        \end{array}
      \right .
\end{align}
and the composition
\begin{align}
  \label{Eq:Numerical Composition Initial Condition}
    C(x,y;t=0) &=
      \left\{
        \begin{array}{ll}
          1 & \text{if} \quad y \ge \frac{1}{2} \, ,\\
          0 & \text{otherwise}                  \, .
        \end{array}
      \right .
\end{align}
We impose boundary conditions on the velocity and temperature as in 
equations~\eqref{Eq:No-FLow BCs}--\eqref{Eq:Free-Slip BCs}
and~\eqref{Eq:Temperature BCs y eq 0}--\eqref{Eq:Temperature BCs x eq 3}, respectively.

All four methods for modeling the location of the compositional boundary are implemented in the 
open source mantle convection code
\href{https://geodynamics.org/cig/software/aspect/}{\textsc{ASPECT}}, which is described in 
detail in \cite{MK-TH-WB:2012} and \cite{TH-JD-RG-WG:2017}.
We use ASPECT to compute the velocity, pressure, and temperature fields associated with the 
underlying flow field.
In other words, the only difference between each of the four methods we use here is the 
specific methodology described in
Sections~\ref{Subsubsection:Finite Element Advection Algorithm with Entropy Viscosity}--\ref{Subsubsection:Particles} above.

In all of our computations we use a $(Q_2, Q_1)$ FEM element combination for the numerical 
solution of the Stokes equations~\eqref{Eq:Weak Formulation of Stokes Equations}--\eqref{Eq:Weak Formulation of the Continuity Equation},
and a $Q_2$ FEM element for the numerical solution of the temperature 
equation~\eqref{Eq:Weak Form of the Discretization of the Temperature Equation}.
We use a second-order accurate spatial discretization for the composition equation with 
FEM-EV and DGBP, a $Q_1$ discontinuous element with the VOF algorithm, and a piecewise 
constant `composition' element for the particle method with an arithmetic cell averaging 
interpolation algorithm. 
See~~\cite{MK-TH-WB:2012} for additional details concerning the 
implementation of the FEM-EV algorithm in ASPECT, \cite{YH-EGP-MIB:2016} for additional 
details concerning the  the implementation of the DGBP algorithm in ASPECT, 
and~\cite{RG-EH-EGP-WB:2016} for additional details concerning the implementation of the 
particle algorithm in ASPECT.

For all of the computations shown here we use a fixed uniform grid with $192 \times 64$ square 
cells each with side $h= 64^{-1}$.
We have also computed the same problems on a uniform grid with $h = 32^{-1}$; i.e., with 
$96 \times 32$ grid cells. 
The computational results on the coarser $96 \times 32$ grid are quite similar to those on 
the finer $192 \times 64$ grid, albeit at a lower resolution.
We have thus determined that our computations on a uniform $192 \times 64$ grid are 
sufficiently well-resolved and accurate to allow us to arrive at the conclusions we discuss 
below.

Note that since we are using a second-order accurate (i.e., $O \left(h^2\right)$)
$(Q_2, Q_1)$ FEM element combination for the velocity, temperature and pressure, our grid 
resolution of $h = 64^{-1}$ \textit{roughly} corresponds to a grid resolution of 
$h= 1 / 4096$ for a first-order accurate $(Q_1, Q_0)$ FEM element combination that is often 
used by researchers to approximate solutions of
equations~\eqref{Eq:Nondimensional Vector Form of the Stokes Equations}--\eqref{Eq:Nondimensional Vector Form of the Temperature Equation} that govern
incompressible convection in the Earth's mantle.

\subsection{Computations at $\mathrm{Ra} = 10^5$ with $\mathrm{B} = 1.0$ fixed}
\label{Subsection:Computations at Ra eq 1e5 with B eq 1 fixed}

Our first set of computations are for a value of $\mathrm{B}$ for which at $\mathrm{Ra} = 10^5$
the flow is very strongly stratified; namely, $\mathrm{B} = 1.0$.
In this regime it is possible for us to use all four algorithms for modeling the location 
of the compositional  field.
As we will see in Section~\ref{Subsection:Computations at Ra eq 1e5 with 0 le B le 1} below, 
this is not true in the regime in which the flow overturns and kinematic mixing occurs; i.e., 
for $\mathrm{B} > \mathrm{B}_{cr}$ where $\mathrm{B}_{cr}$ is,  for $\mathrm{Ra} = 10^5$, the
value of $\mathrm{B}$ at which the compositional density fields overturn and kinematic mixing 
begins to occur.

\begin{figure}[tbph!]
    \centering
    \includegraphics[width=1.0\linewidth]{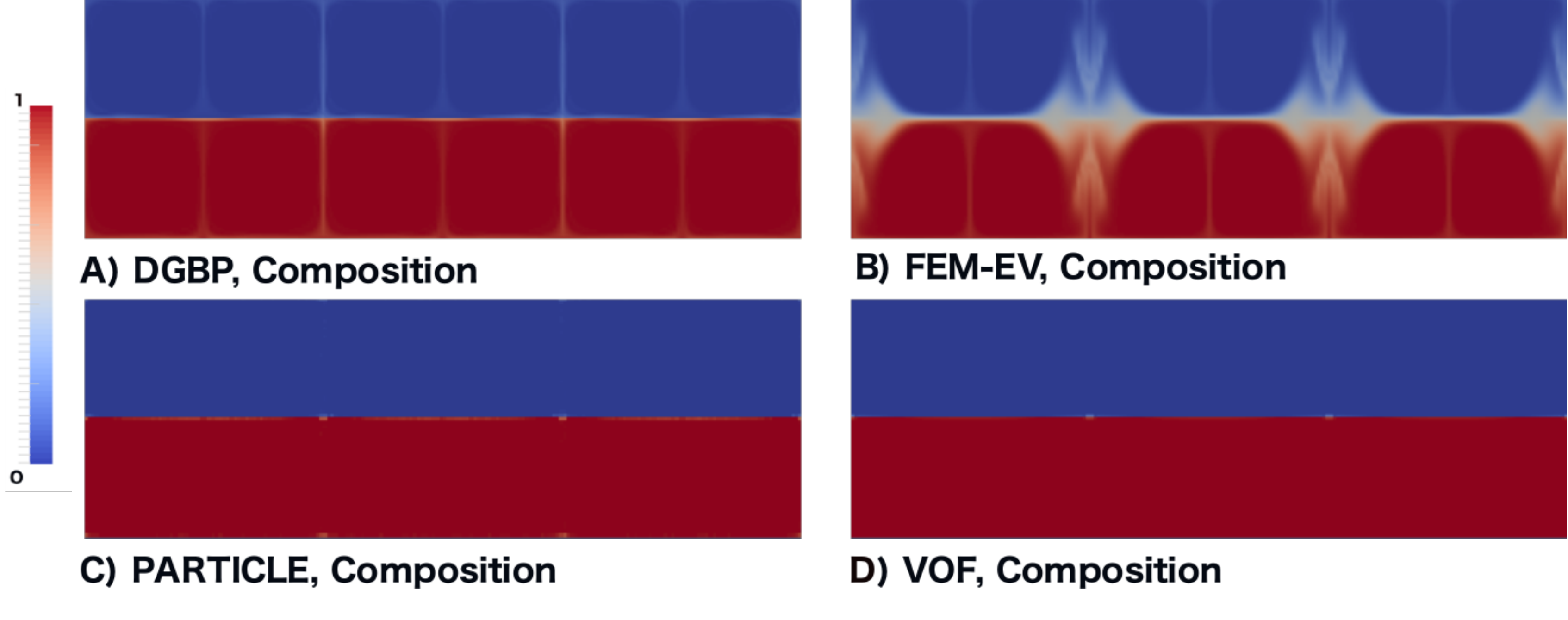} 
    \caption{The composition at $t' = 0.05$ for $\mathrm{Ra} = 10^5$ and $B=1.0$.
    }
    \label{Fig:COMPOSITION_Combined_t_005}
\end{figure}

\begin{figure}[tbph!]
    \centering
    \includegraphics[width=1.0\linewidth]{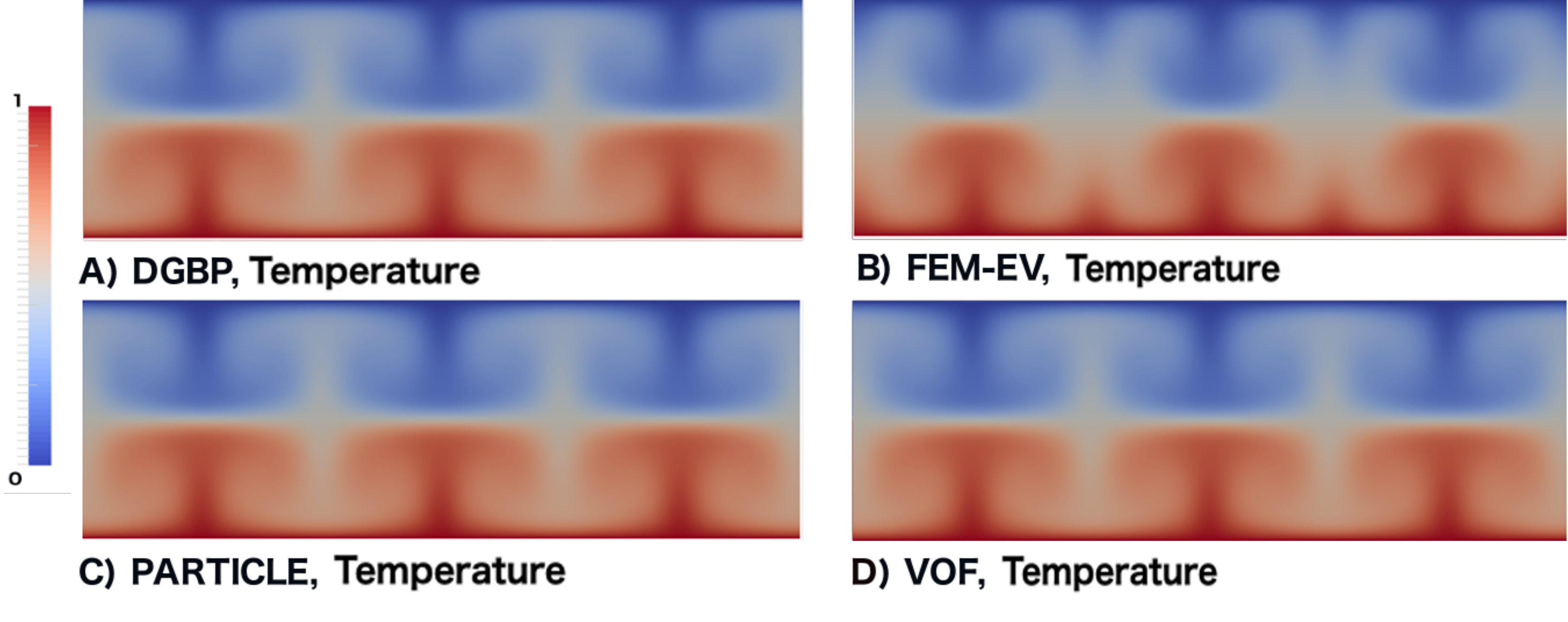} 
    \caption{The temperature at $t' = 0.05$ for $\mathrm{Ra} = 10^5$ and $B=1.0$.}
    \label{Fig:TEMPERATURE_Combined_t_005}
\end{figure}

\begin{figure}[tbph!]
    \centering
    \includegraphics[width=1.0\linewidth]{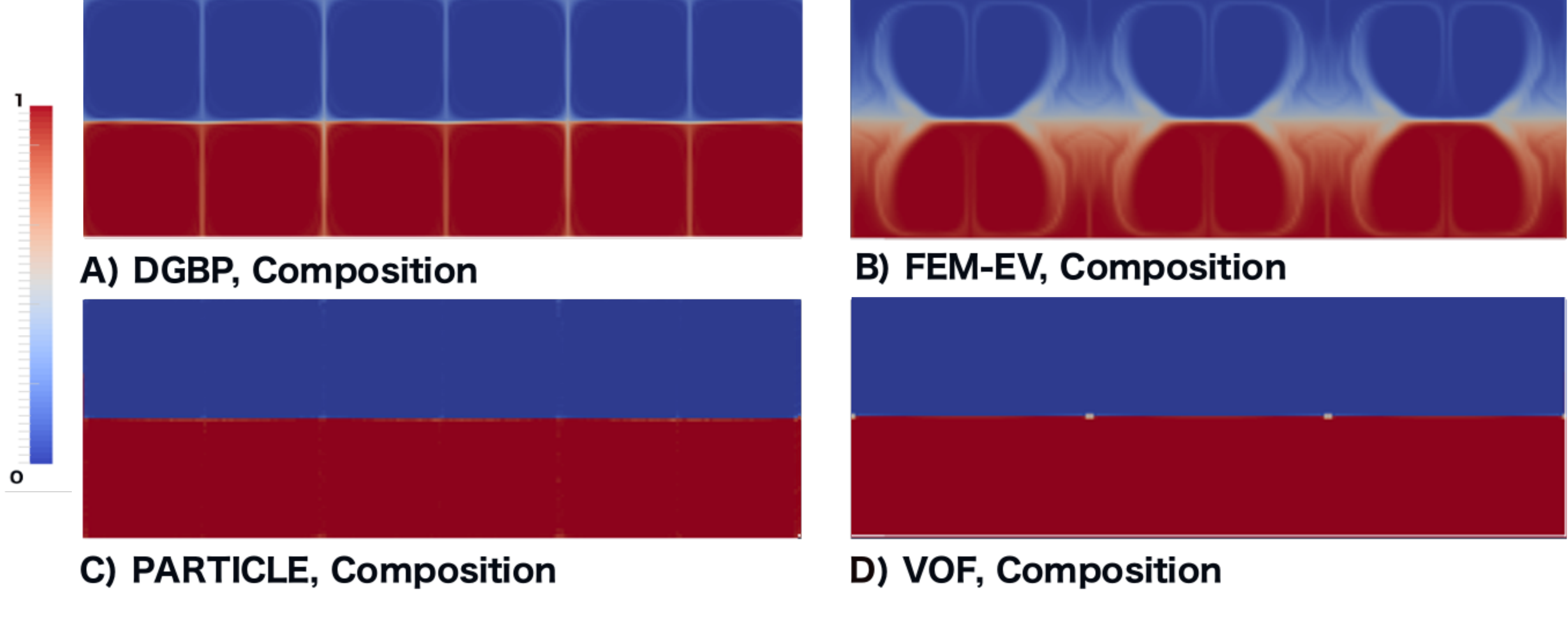} 
    \caption{The composition at $t' = 0.1$ for $\mathrm{Ra} = 10^5$ and $B=1.0$.}
    \label{Fig:COMPOSITION_Combined_t_010}
\end{figure}

\begin{figure}[tbph!]
    \centering
    \includegraphics[width=1.0\linewidth]{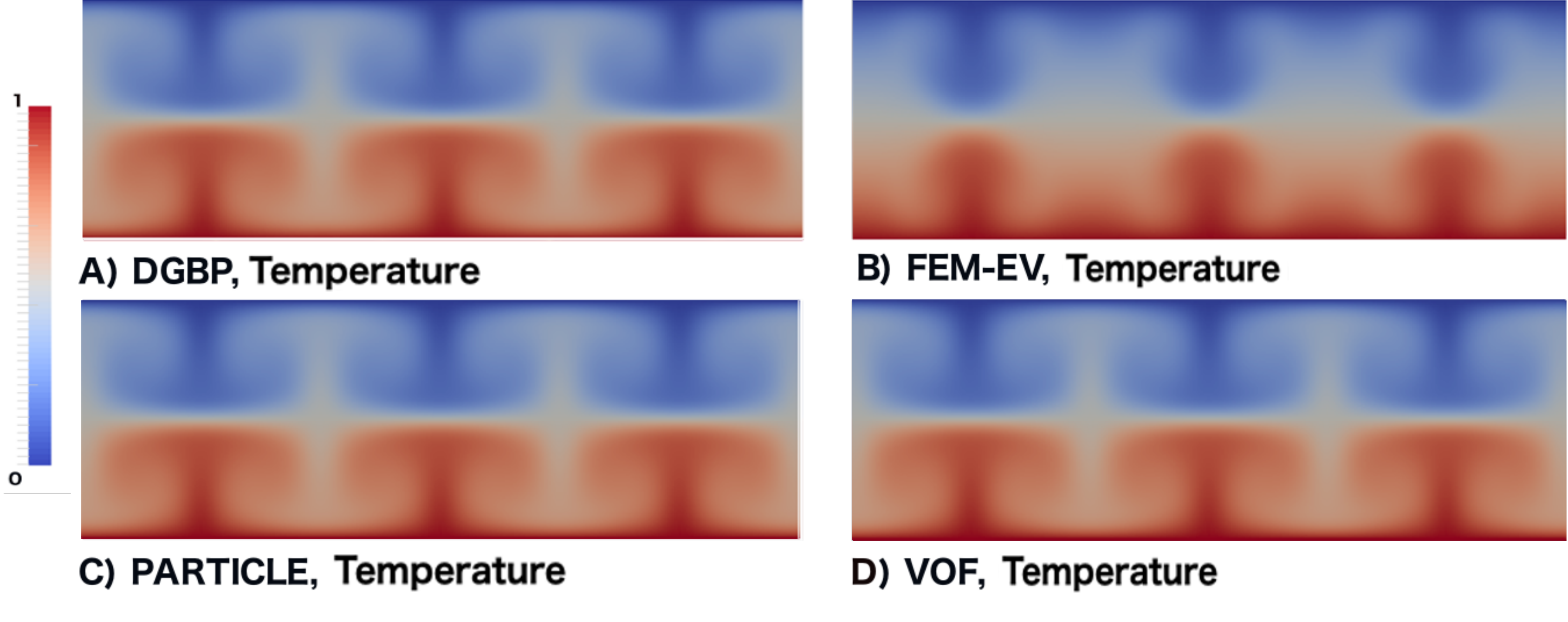} 
    \caption{The temperature at $t' = 0.1$ for $\mathrm{Ra} = 10^5$ and $B=1.0$.}
    \label{Fig:TEMPERATURE_Combined_t_010}
\end{figure}

\begin{figure}[tbph!]
    \centering
    \includegraphics[width=1.0\linewidth]{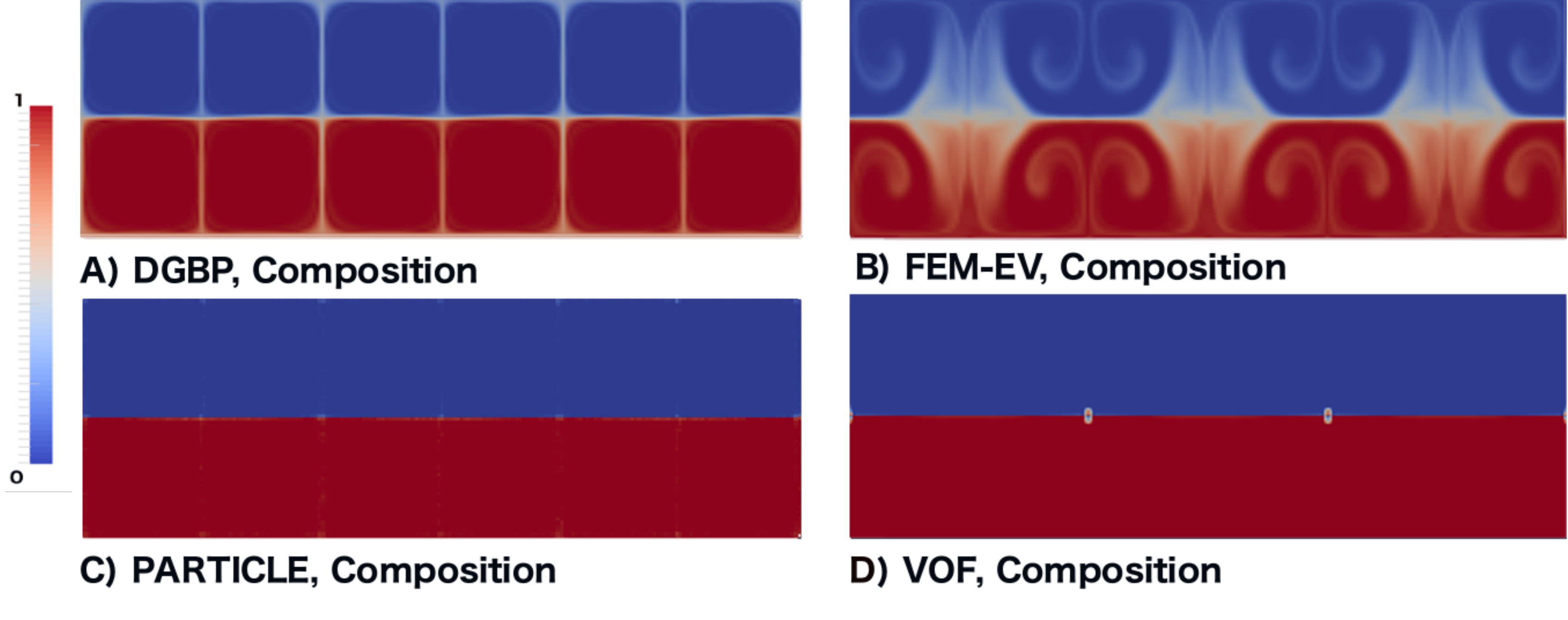} 
    \caption{The composition at $t' = 0.15$  for $\mathrm{Ra} = 10^5$ and $B=1.0$.}
    \label{Fig:COMPOSITION_Combined_t_015}
\end{figure}

\begin{figure}[tbph!]
    \centering
    \includegraphics[width=1.0\linewidth]{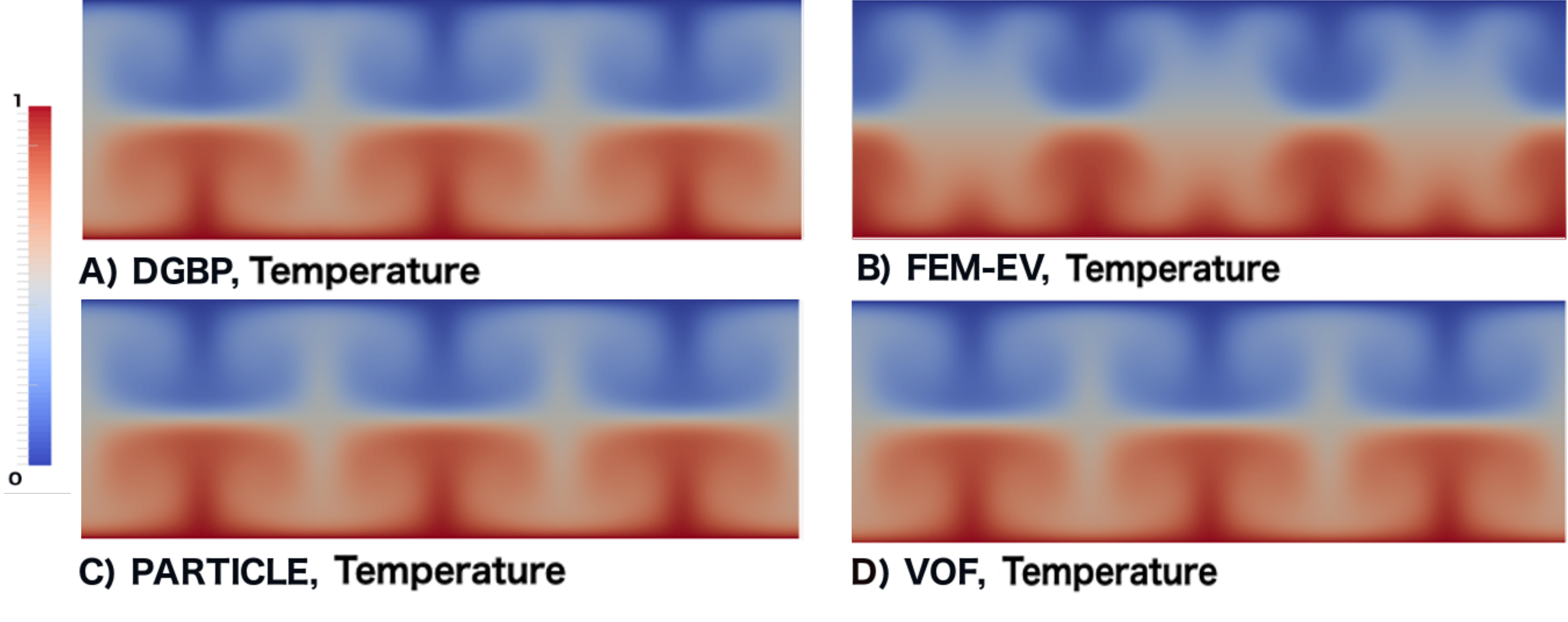} 
    \caption{The temperature at $t' = 0.15$ for $\mathrm{Ra} = 10^5$ and $B=1.0$.}
    \label{Fig:TEMPERATURE_Combined_t_015}
\end{figure}

For each of the four algorithms we show the computational results for the composition field and the 
temperature at dimensionless times
$t^\prime =0.05$,  $t^{\prime}=0.10$, and $t^\prime = 0.15$.
These are displayed in Figures~\ref{Fig:COMPOSITION_Combined_t_005} 
and~\ref{Fig:TEMPERATURE_Combined_t_005}, Figures ~\ref{Fig:COMPOSITION_Combined_t_010} 
and~\ref{Fig:TEMPERATURE_Combined_t_010}, and Figures~\ref{Fig:COMPOSITION_Combined_t_015} 
and~\ref{Fig:TEMPERATURE_Combined_t_015}, respectively.

\subsection{The performance of the four algorithms for $\mathrm{Ra} = 10^5$ and $\mathrm{B} = 1.0$}
\label{Subsection:The performance of the four algorithms for Ra = 1e5 and B eq 1.0}

In the following discussion we will use the following notation.
We divide the computational domain $\Omega$ into two distinct regions
$\Omega = \Omega_1 \cup \Omega_2$ where $\Omega_1 = \{ (x,y) : y > 0.5 \}$ denotes the upper half of the domain and $\Omega_2 = \{ (x,y) : y < 0.5 \}$ denotes the lower half of the domain.

\subsubsection{The performance of the DGBP algorithm}
\label{Subsubsection:The performance of the DGBP algorithm}

Figures~\ref{Fig:COMPOSITION_Combined_t_005}--\ref{Fig:TEMPERATURE_Combined_t_015} demonstrate 
that, at each time, the composition and temperature we computed with the DGBP, VOF and Particle algorithms are quite similar to each other, especially the temperature field.
The flow is organized into relatively steady-state, discrete cells.
Furthermore, the temperature fields we computed with these three algorithms are virtually 
indistinguishable from one another at each of the times shown.
Note, however, that the DGBP algorithm has visible white bands at the interface between each 
convection cells. 
These are locations where the values of the composition field is approximately $C=0.5$ indicating 
that some of the composition field with value $C = 1$ that was initially in the subdomain 
$\Omega_1$ has been entrained into the the flow in $\Omega_2$ and vice-versa.
Although this enables one to easily see the convection cells it is due to numerical error.

\begin{figure}[b!]
    \begin{subfigure}{.45\textwidth}
        \includegraphics[width=1.0\linewidth]{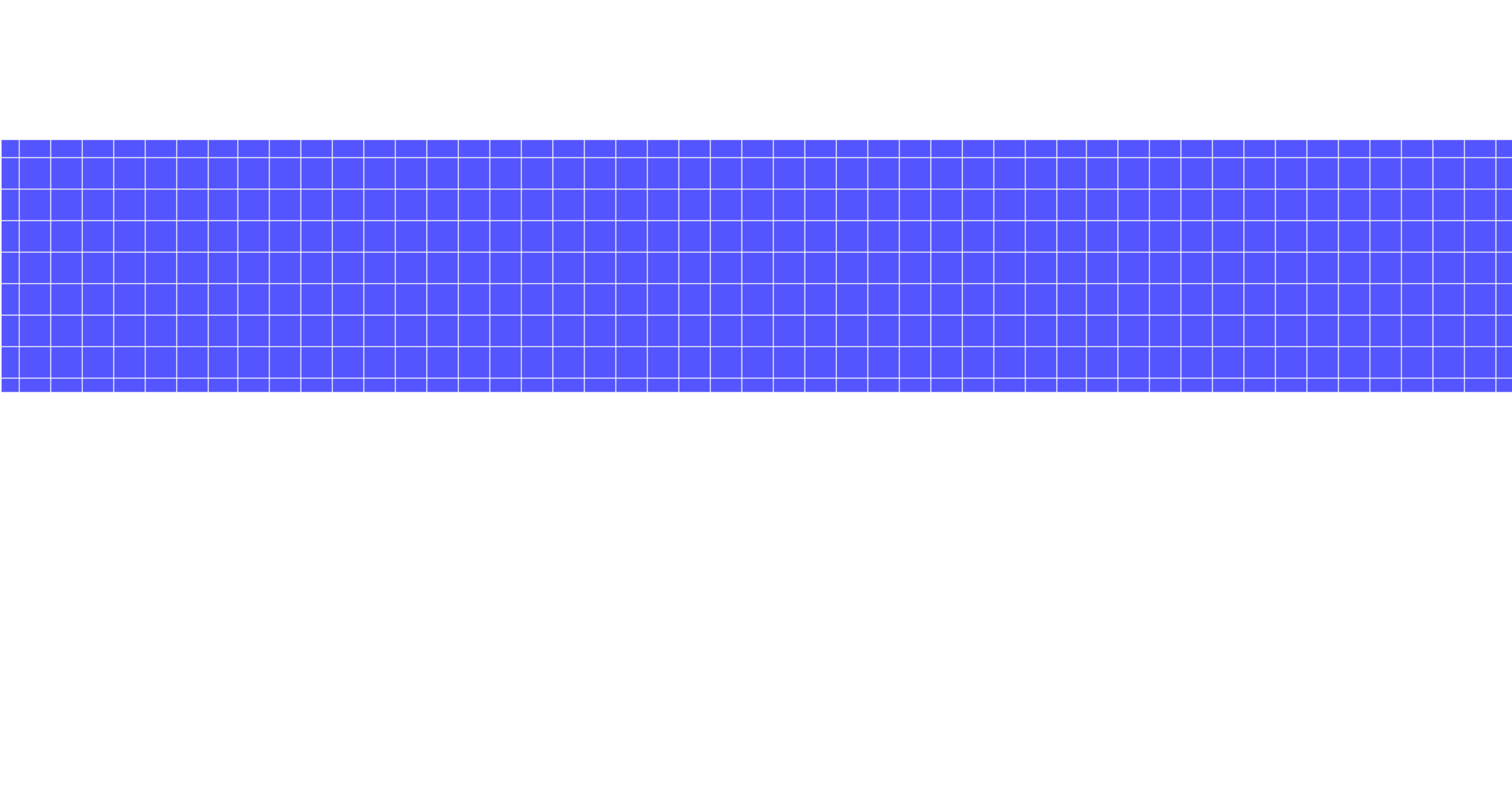} 
        \caption{\small Particles initially in $\Omega_1$ at $t'=0$}
    \end{subfigure}
    \begin{subfigure}{.45\textwidth}
        \includegraphics[width=1.0\linewidth]{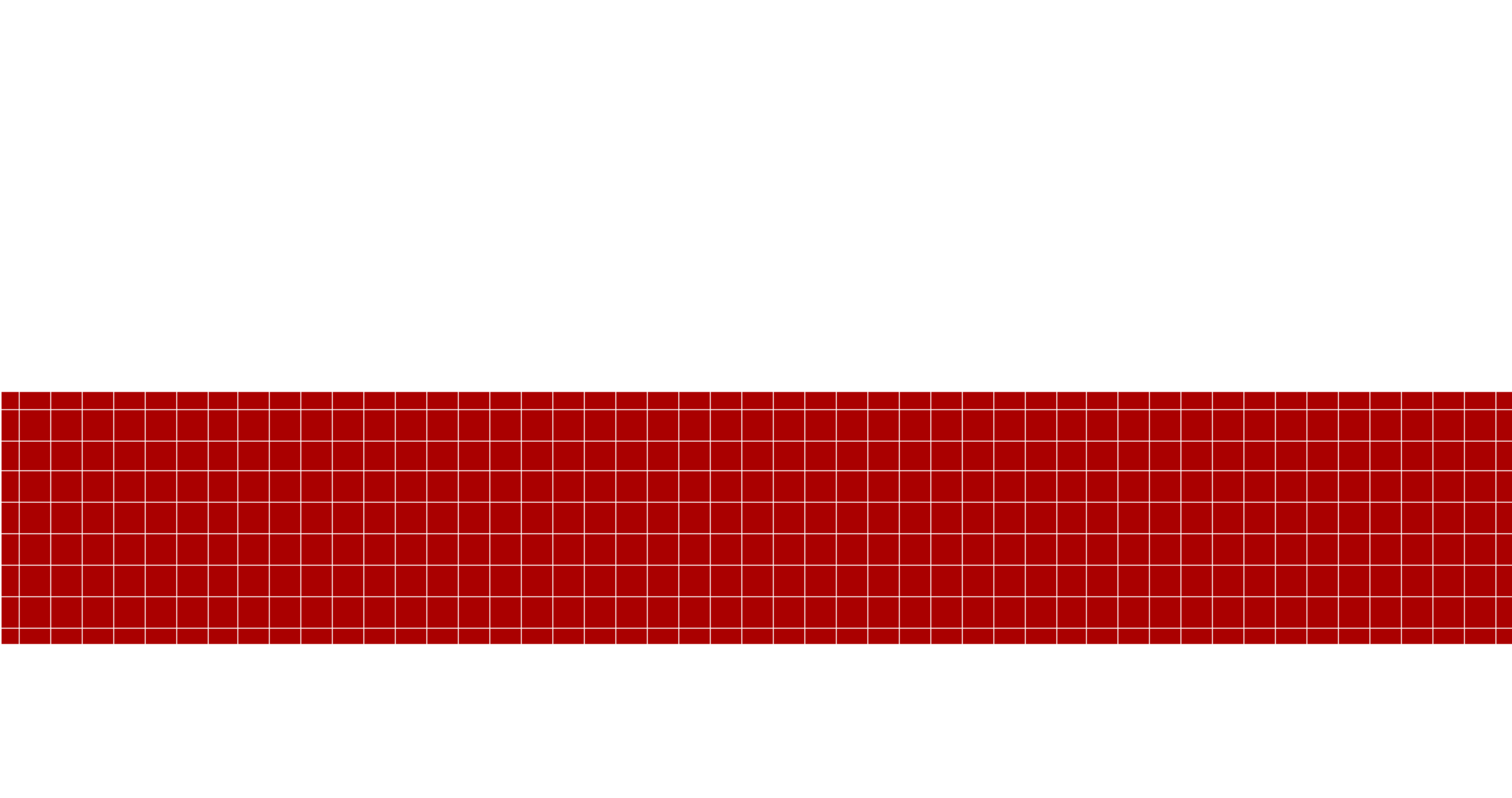}
        \caption{\small  Particles initially in $\Omega_2$ at $t'=0$}
    \end{subfigure}
    \begin{subfigure}{0.45\textwidth}
        \includegraphics[width=1.0\linewidth]{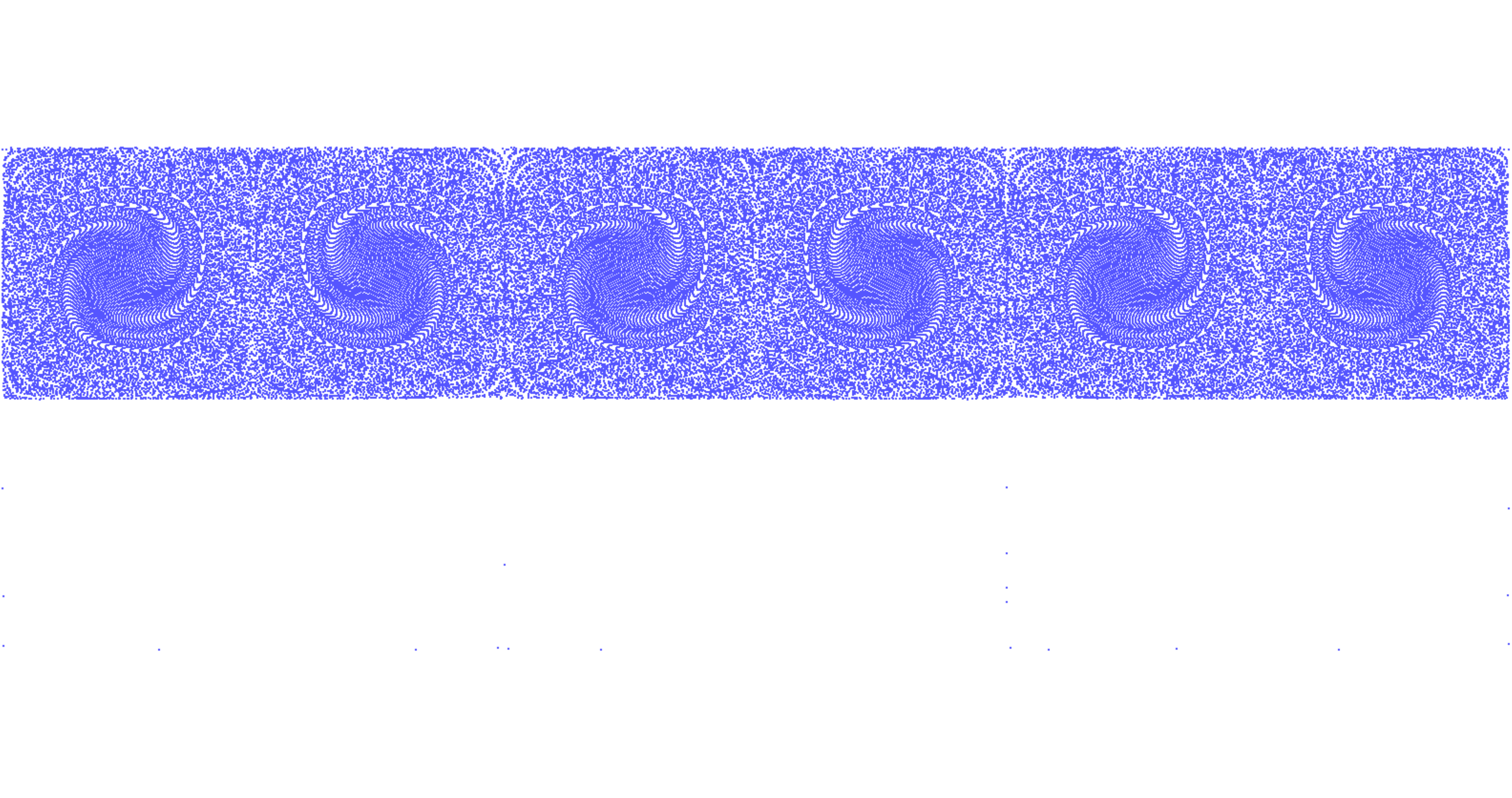} 
        \caption{\small Particles from $\Omega_1$ at $t'=0.05$}
    \end{subfigure}
    \begin{subfigure}{0.45\textwidth}
        \includegraphics[width=1.0\linewidth]{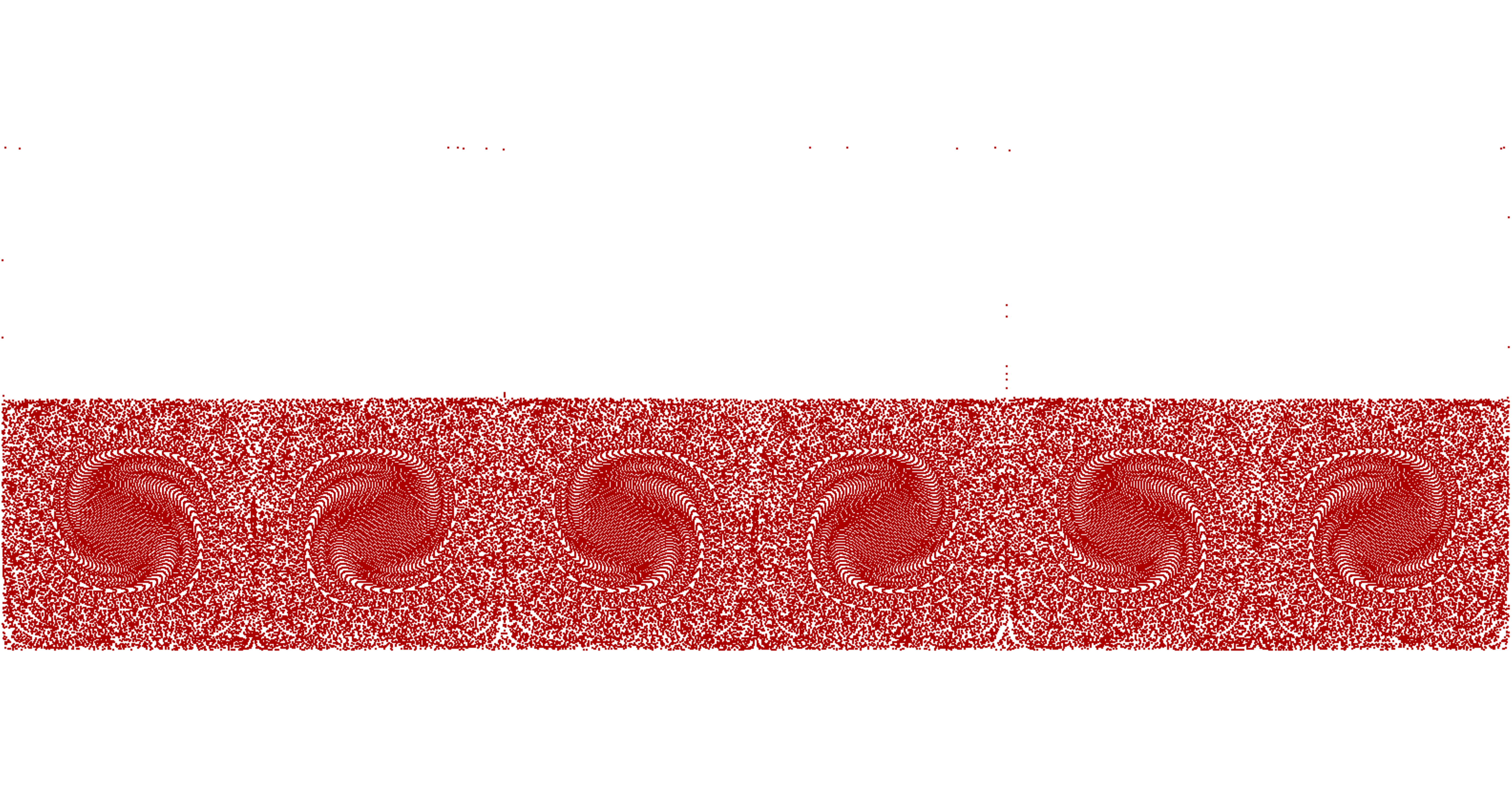}
        \caption{\small Particles from $\Omega_2$ at $t'=0.05$}
    \end{subfigure}
    \begin{subfigure}{0.45\textwidth}
        \includegraphics[width=1.0\linewidth]{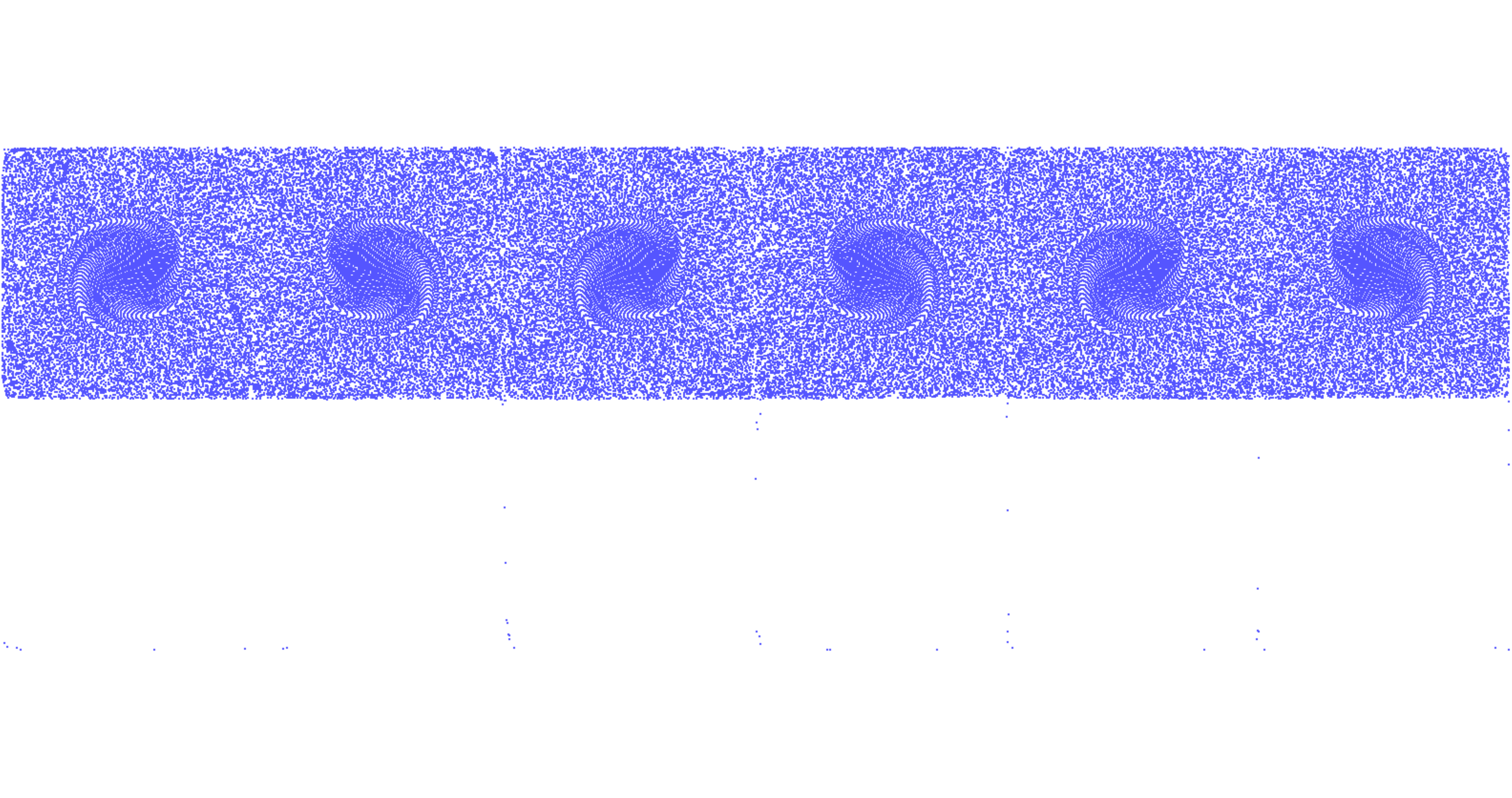} 
        \caption{\small Particles from $\Omega_1$ at $t'=0.10$}
    \end{subfigure}
    \begin{subfigure}{0.45\textwidth}
        \includegraphics[width=1.0\linewidth]{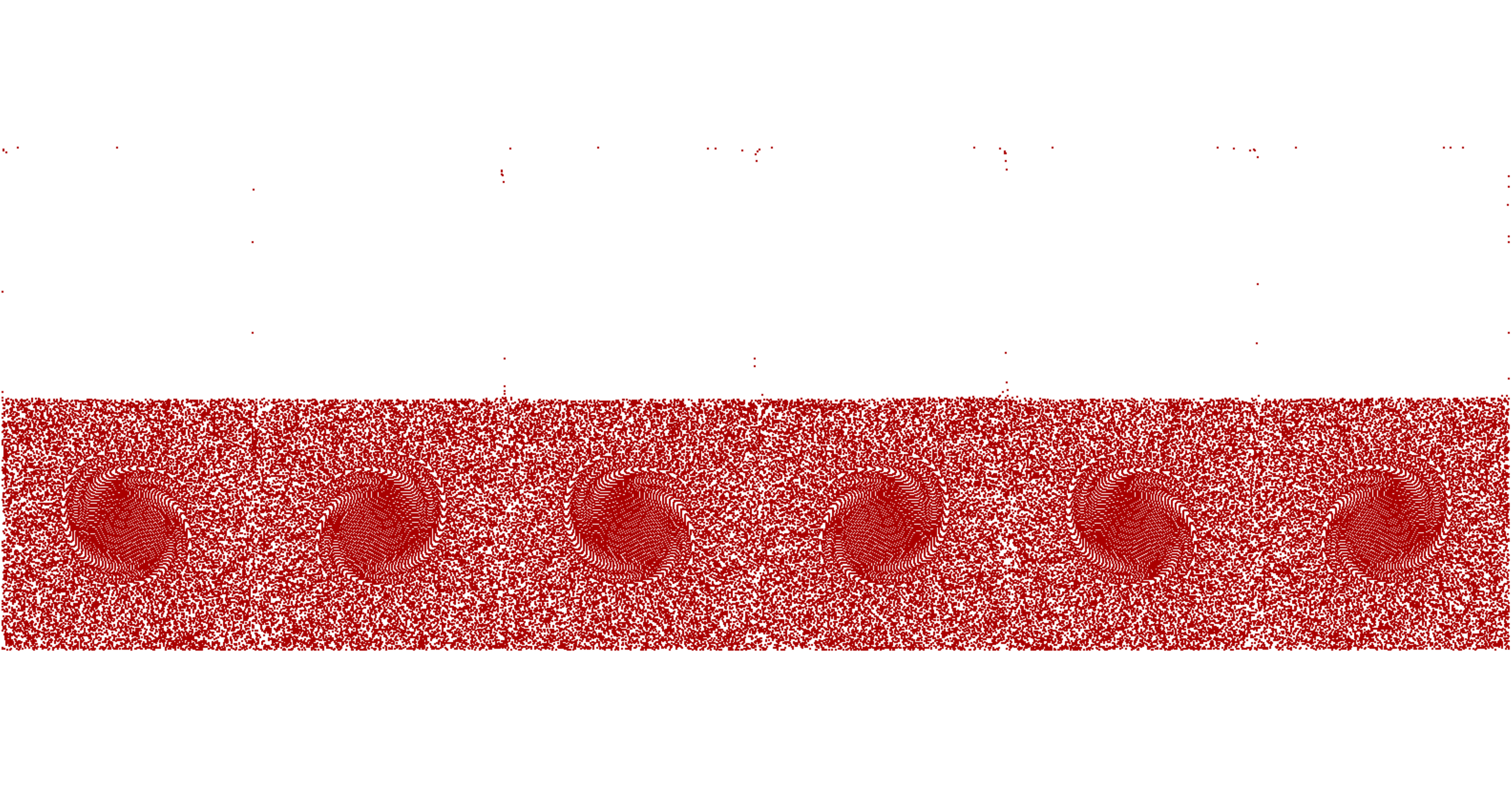}
        \caption{\small Particles from $\Omega_2$ at $t'=0.10$}
    \end{subfigure}
    \begin{subfigure}{0.45\textwidth}
        \includegraphics[width=1.0\linewidth]{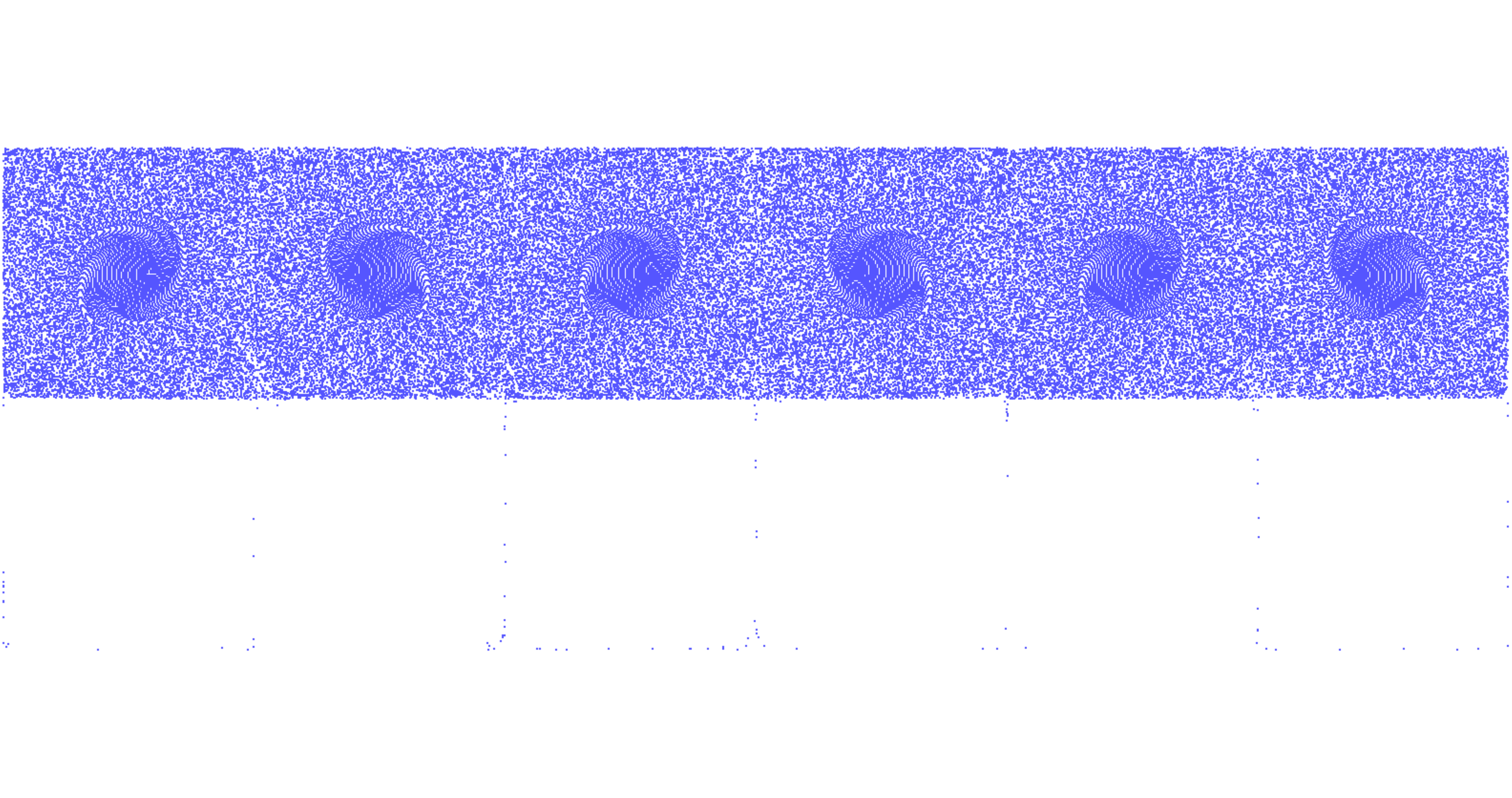} 
        \caption{\small  Particles from $\Omega_1$ at $t'=0.15$}
    \end{subfigure}
    \begin{subfigure}{0.45\textwidth}
        \includegraphics[width=1.0\linewidth]{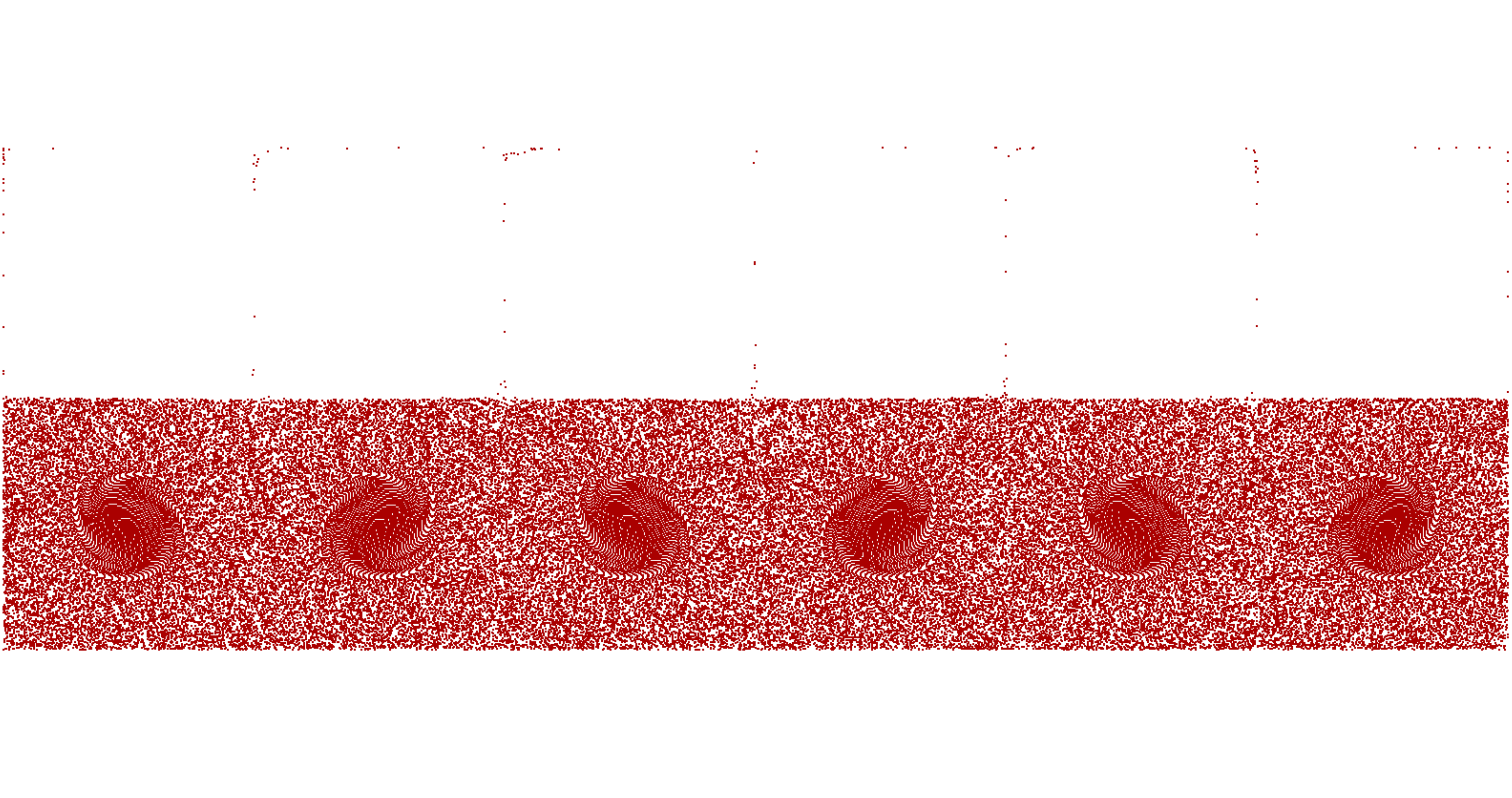}
        \caption{\small Particles from $\Omega_2$ at $t'=0.15$}
    \end{subfigure}
    \centering
    \caption{The particles at times $t' = 0,\,0.05,\,0.10,\,0.15$ for a computation with 
        $\mathrm{Ra} = 10^5$ and $\mathrm{B} = 1$ on a $192 \times 64$ uniform grid.
        Initially the particles are distributed uniformly within each cell with
        $4 \times 4 = 16$ particles per cell, so there are a total of 196,608 particles 
        in the computation.
        Note that for $t' > 0.0$ some blue particles that should remain in
        $\Omega_1 = \left[0, 0.5\right] \times \left[0, 3\right]$ are entrained in the
        flow along the boundaries of the convection cells and move into
        $\Omega_2 = \left[0.5 , 1.0\right] \times \left[0, 3\right]$ and vice-versa.
        See Table~\ref{Table:Number of Particles Entrained with Initially 16 per Cell} for 
        additional details.              
    }
    \label{Fig:16 Particles per Cell}
\end{figure}
\begin{figure}[t!]
    \begin{subfigure}{.45\textwidth}
        \includegraphics[width=1.0\linewidth]{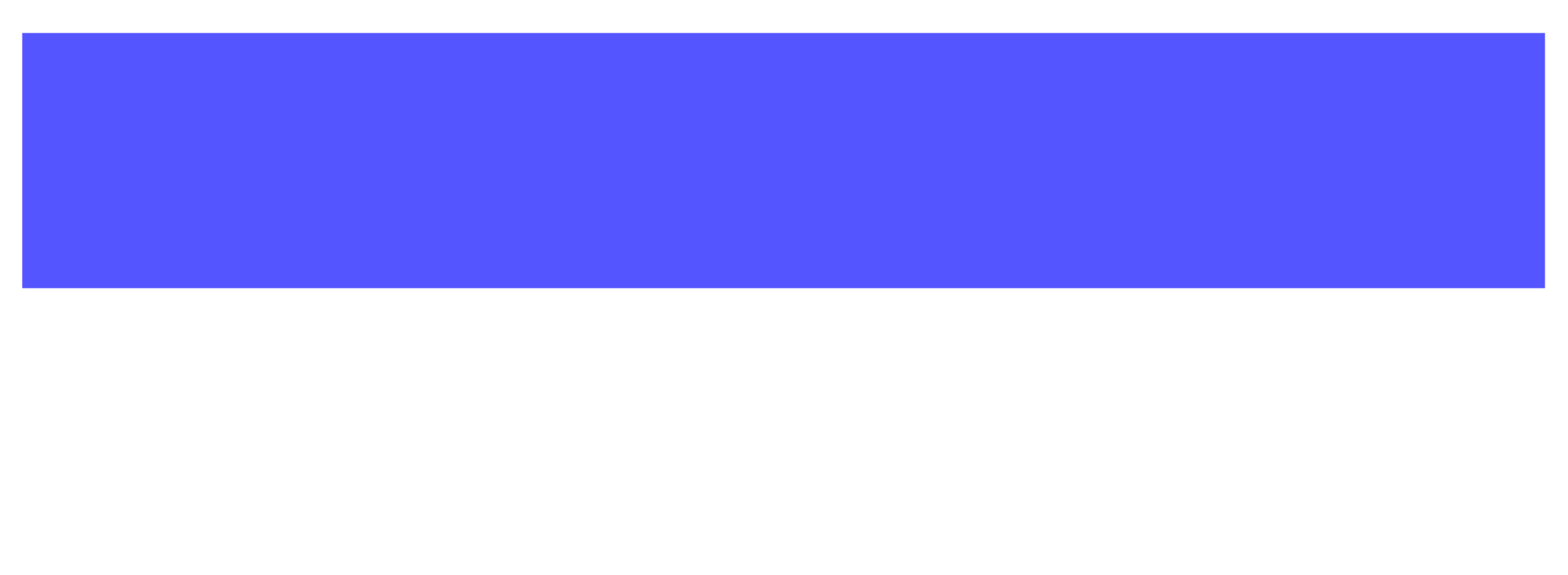} 
        \caption{\small Particles initially in $\Omega_1$ at $t'=0$}
    \end{subfigure}
    \begin{subfigure}{.45\textwidth}
        \includegraphics[width=1.0\linewidth]{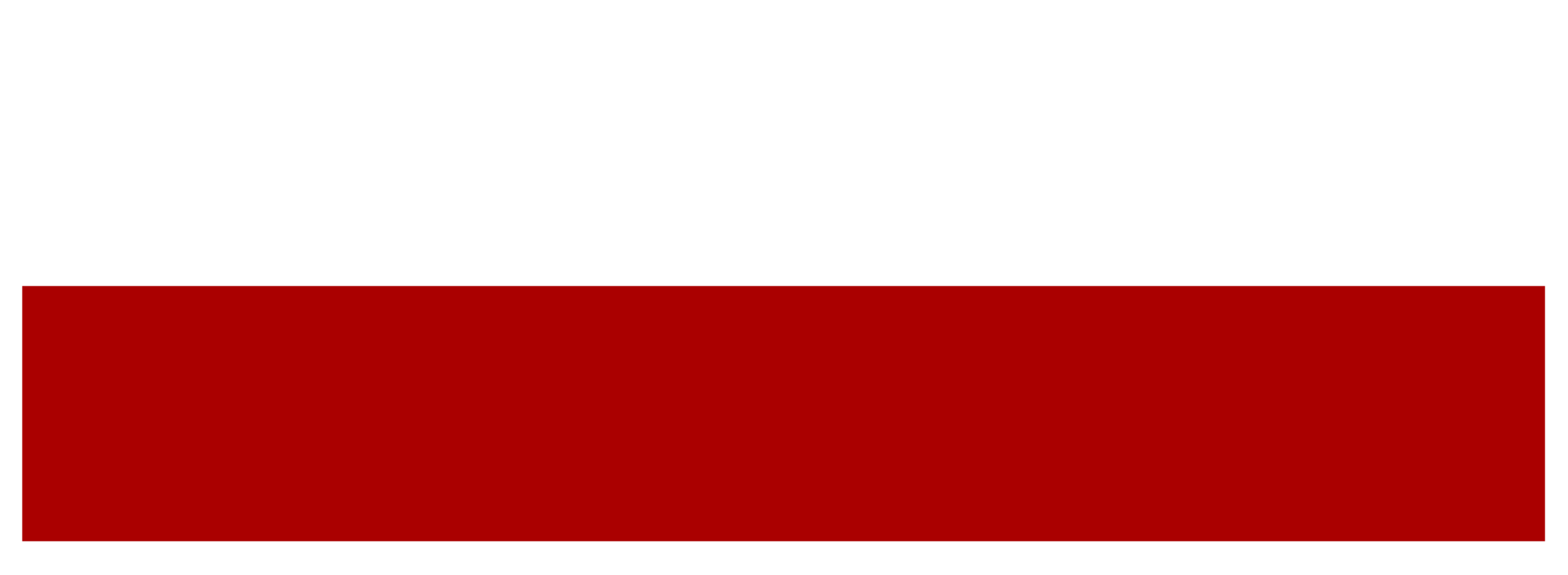}
        \caption{\small Particles initially in $\Omega_2$ at $t'=0$}
    \end{subfigure}
    \begin{subfigure}{0.45\textwidth}
        \includegraphics[width=1.0\linewidth]{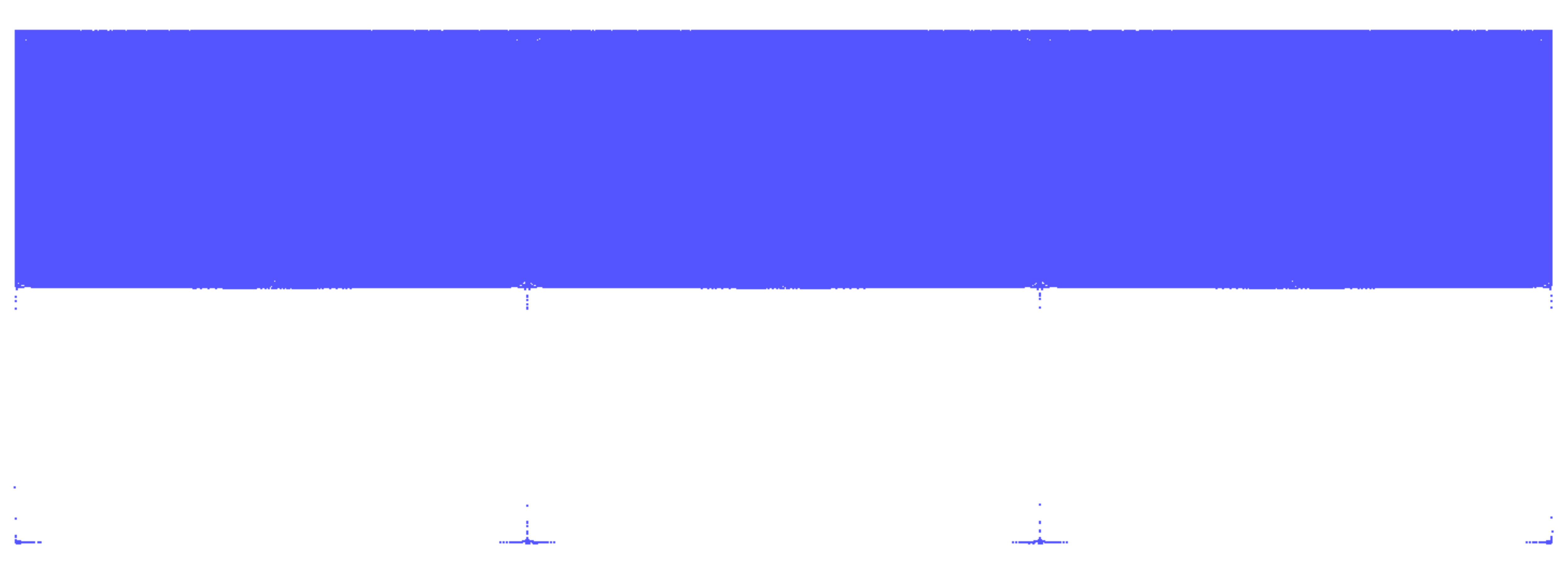} 
        \caption{\small Particles from $\Omega_1$ at $t'=0.05$}
    \end{subfigure}
    \begin{subfigure}{0.45\textwidth}
        \includegraphics[width=1.0\linewidth]{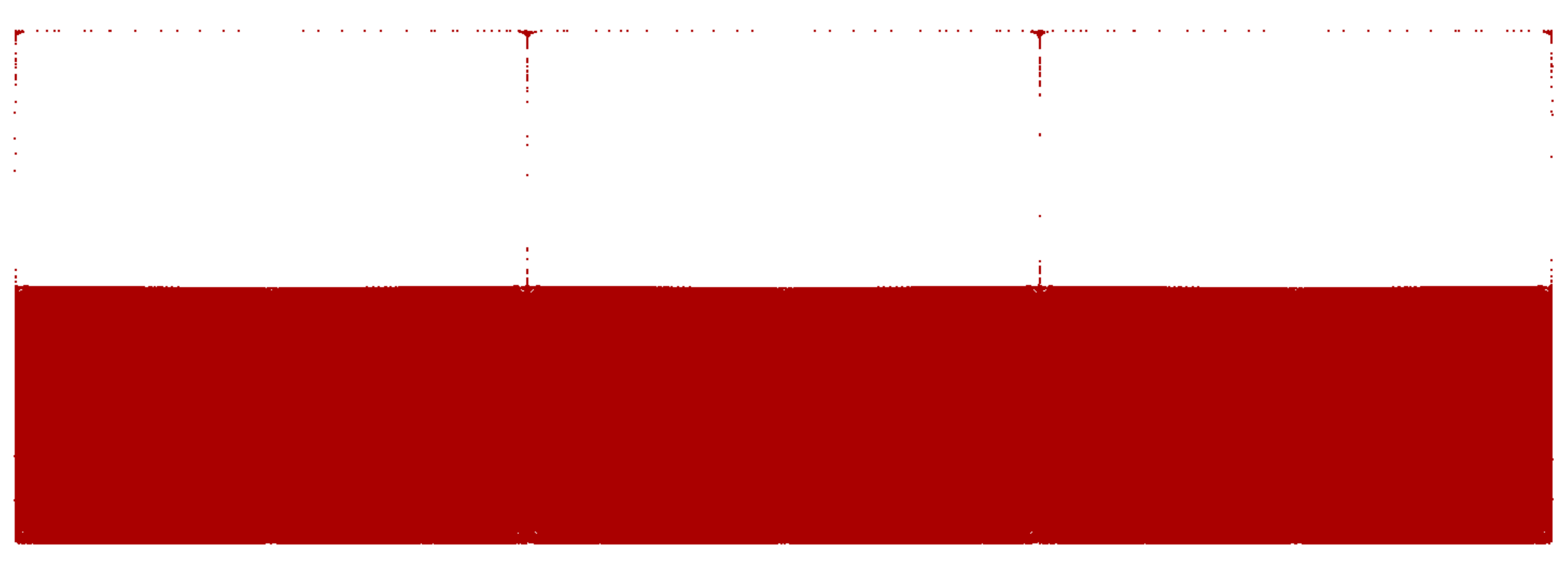}
        \caption{\small Particles from $\Omega_2$ at $t'=0.05$}
    \end{subfigure}
    \begin{subfigure}{0.45\textwidth}
        \includegraphics[width=1.0\linewidth]{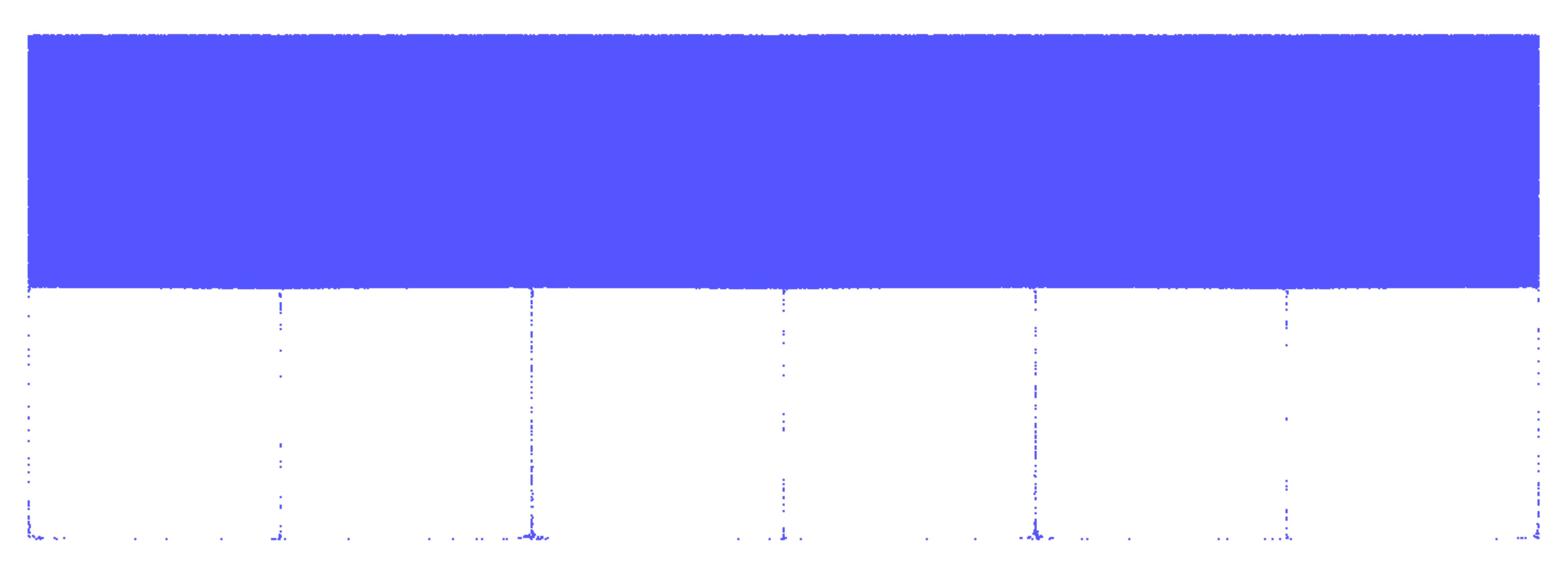} 
        \caption{\small Particles from $\Omega_1$ at $t'=0.10$}
    \end{subfigure}
    \begin{subfigure}{0.45\textwidth}
        \includegraphics[width=1.0\linewidth]{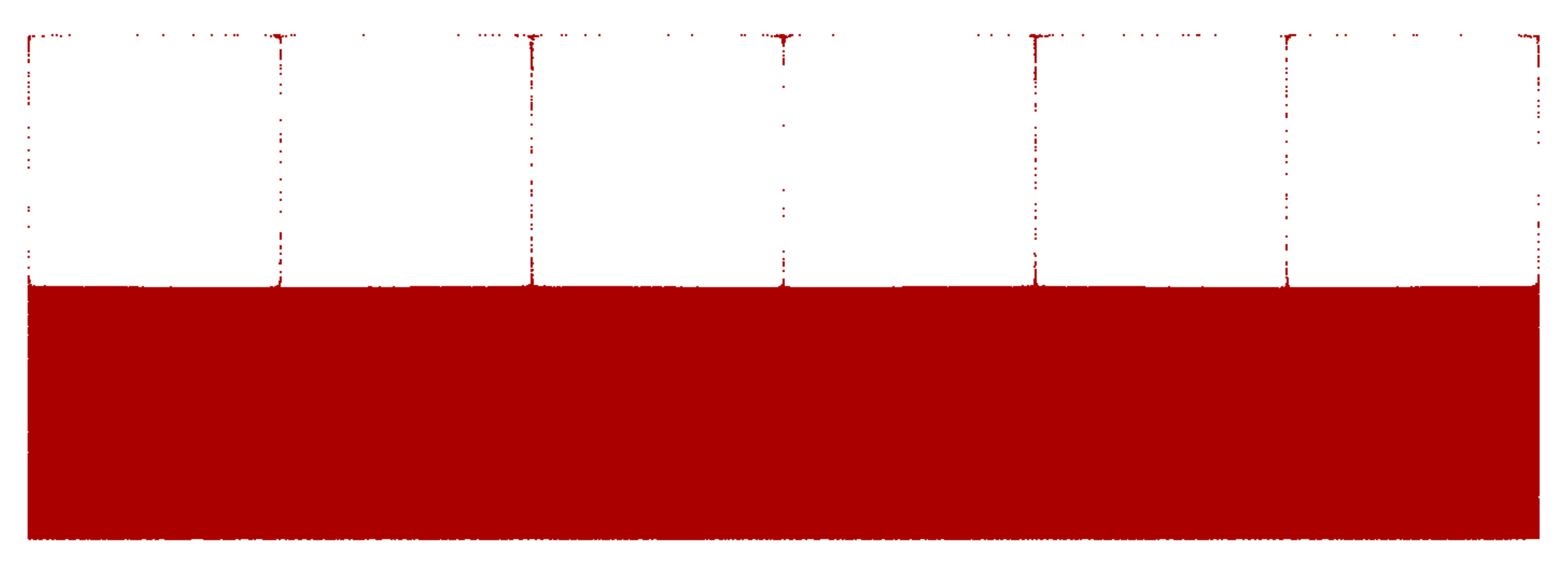}
        \caption{\small Particles from $\Omega_2$ at $t'=0.10$}
    \end{subfigure}
    \begin{subfigure}{0.45\textwidth}
        \includegraphics[width=1.0\linewidth]{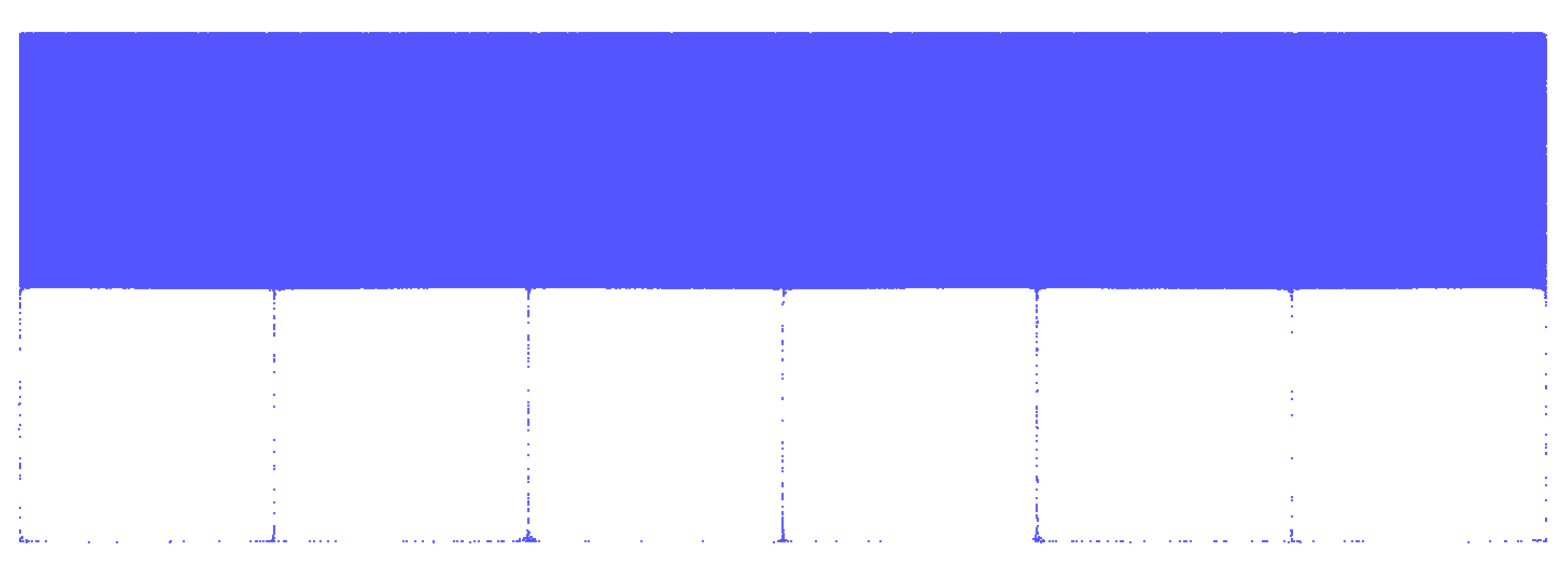} 
        \caption{\small Particles from $\Omega_1$ at $t'=0.15$}
    \end{subfigure}
    \begin{subfigure}{0.45\textwidth}
        \includegraphics[width=1.0\linewidth]{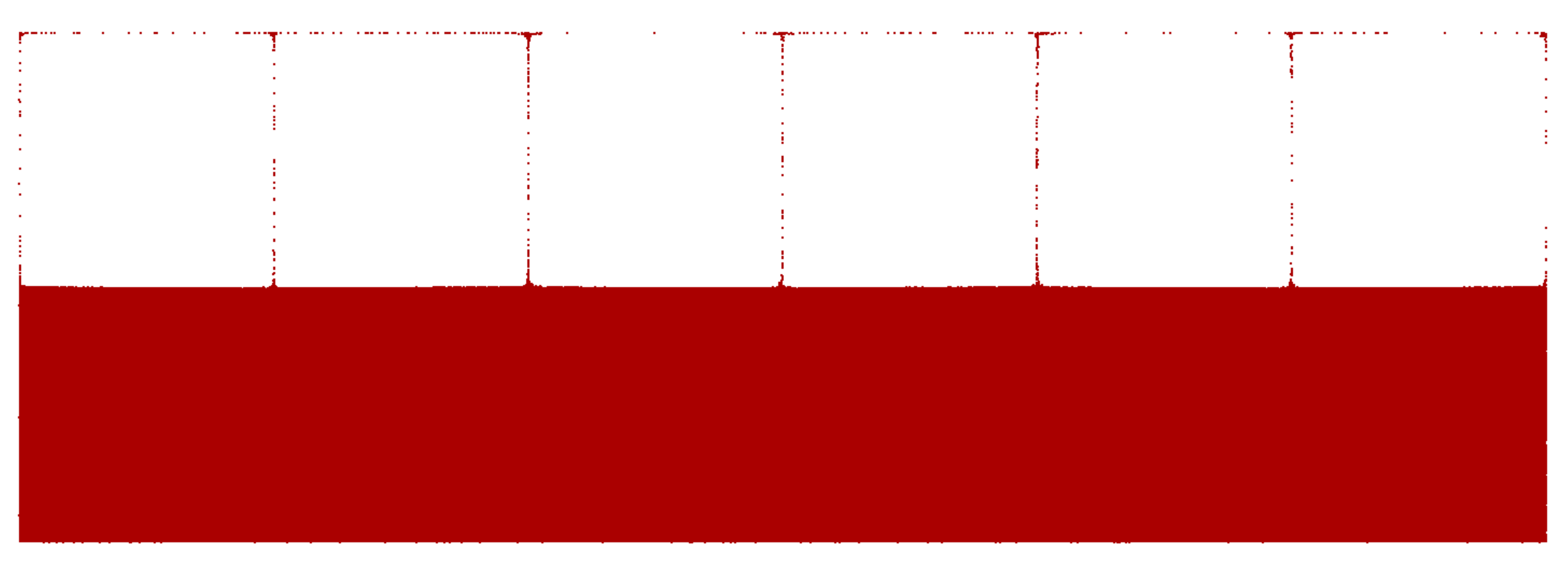}
        \caption{\small Particles from $\Omega_2$ at $t'=0.15$}
    \end{subfigure}
    \centering
    \caption{The particles at times $t' = 0,\,0.05,\,0.10,\,0.15$ for a computations with 
        $\mathrm{Ra} = 10^5$ and $\mathrm{B} = 1$ on a $192 \times 64$ uniform grid.
        Initially the particles are distributed uniformly within each cell with
        $16 \times 16 = 256$ 
        particles per cell, so there are a total of $3,145,728$ particles in the computation.
        Note that for $t' > 0.0$ some blue particles that should remain in
        $\Omega_1 = \left[0, 0.5\right] \times \left[0, 3\right]$ are entrained in the
        flow along the boundaries of the convection cells and move into
        $\Omega_2 = \left[0.5 , 1.0\right] \times \left[0, 3\right]$ and vice-versa.
        See Table~\ref{Table:Number of Particles Entrained with Initially 256 per Cell} for 
        additional details.
    }
    \label{Fig:256 Particles per Cell}
\end{figure}
\begin{table}[b!]
    \centering
    \begin{tabular}{|l|l|c|c|c|}
        \hline
        Time& Direction & \#  entrained particles & Percentage \\
        \hline
        $t'=0.05$   &$\Omega_1\longrightarrow \Omega_2$  & 92  & 0.05\%\\
        &$\Omega_2\longrightarrow \Omega_1$  & 47  & 0.02\%\\
        \hline
        $t'=0.10$   &$\Omega_1\longrightarrow \Omega_2$  & 128 & 0.07\%\\
        &$\Omega_2\longrightarrow \Omega_1$  & 117 & 0.06\%\\
        \hline
        $t'=0.15$   &$\Omega_1\longrightarrow \Omega_2$  & 176 & 0.09\%\\
        &$\Omega_2\longrightarrow \Omega_1$  & 173 & 0.09\%\\
        \hline
    \end{tabular}   
    \caption{The number particles from each subdomain $\Omega_1$, $\Omega_2$, that have been 
        entrained  and advected into the other subdomain.
        This computation is on a uniform $192 \times 64$ grid with $\mathrm{Ra} = 10^5$ and 
        $\mathrm{B= 1}$ and, initially, $16$ particles per cell for a total of 196,608 particles.
    } 
    \label{Table:Number of Particles Entrained with Initially 16 per Cell}. 
\end{table}

\subsubsection{The performance of the particle algorithm}
\label{Subsubsection:The performance of the particle algorithm}

In Figures~\ref{Fig:16 Particles per Cell} and~\ref{Fig:256 Particles per Cell} below the particles 
initially in $\Omega_1$ are blue and particles initially in $\Omega_2$ are red.
It is difficult to see in Figures~\ref{Fig:COMPOSITION_Combined_t_005}, 
~\ref{Fig:COMPOSITION_Combined_t_010}, and~\ref{Fig:COMPOSITION_Combined_t_015},
but in our computations some particles from $\Omega_1$ have been entrained and advected downward 
into $\Omega_2$, the red region.  
Similarly, some particles from $\Omega_2$ have been entrained and advected upward into  from  $\Omega_1$.
This is easier to see in Figures~\ref{Fig:16 Particles per Cell}
and~\ref{Fig:256 Particles per Cell}.

Since the physical model is strongly stratified both the DGBP and particle algorithms are 
exhibiting non-physical features due to numerical errors intrinsic to the algorithm. 
As the time increases, this error increases; the white boundaries shown in the DGBP results 
increase  with time and the number of particles in the wrong domain increases with time
as shown in Tables~\ref{Table:Number of Particles Entrained with Initially 16 per Cell}
and~\ref{Table:Number of Particles Entrained with Initially 256 per Cell}.

\begin{table}[h!]
    \centering
    \begin{tabular}{|l|l|c|c|}
        \hline
        Time& Direction & \#  entrained particles & Percentage \\
        \hline
        $t'=0.05$   &$\Omega_1\longrightarrow \Omega_2$  & 2412   & 0.05\%\\
        &$\Omega_2\longrightarrow \Omega_1$  & 764    & 0.02\%\\
        \hline
        $t'=0.10$   &$\Omega_1\longrightarrow \Omega_2$  & 2648   & 0.08\%\\
        &$\Omega_2\longrightarrow \Omega_1$  & 947    & 0.03\%\\
        \hline
        $t'=0.15$   &$\Omega_1\longrightarrow \Omega_2$  & 2941   & 0.09\%\\
        &$\Omega_2\longrightarrow \Omega_1$  & 1200   & 0.04\%\\
        \hline
    \end{tabular}
    \caption{The number particles from each subdomain $\Omega_1$, $\Omega_2$, that have been 
        entrained and advected into the other subdomain.
        This computation is on a uniform $192 \times 64$ grid with $\mathrm{Ra} = 10^5$ and 
        $\mathrm{B= 1}$ and initially $256$ particles per cell for a total of 3,145,728 particles.}
    \label{Table:Number of Particles Entrained with Initially 256 per Cell}
\end{table}

\subsubsection{The performance of the VOF method}
\label{Subsubsection:The performance of the VOF method}

In contrast, since it is an interface tracking method, the VOF algorithm maintains a sharp 
boundary at $y = 0.5$.
There are two small regions located at each intersection of the corners of two counter 
rotating convection cells.
this is an area with a classic shear flow centered about a stagnation point that lies precisely 
at each corner. 
This is a very difficult flow in which to maintain a well defined interface. 
In future work we will examine the quality of the numerical solution at this point under grid 
resolution.
In particular, since the VOF algorithm is only tracking a set of co-dimension one in a two 
dimensional flow (i.e, a one-dimensional curve in a two dimensional flow) this problem 
is an ideal candidate for adaptive mesh refinement.
One of the key features of ASPECT is it's adaptive mesh refinement capability.
We have used it in other VOF computations to refine only about the interface.
However, we decided that to do so here would be beyond the scope of this paper.
 
\subsubsection{The performance of the FEM-EV algorithm}
\label{Subsubsection:The performance of the FEM-EV algorithm}
 
In stark contrast with the results of the three algorithms discussed above, the results 
for both the composition and temperature fields we computed with the FEM-EV algorithm are
completely dissimilar from those computed with the other three algorithms at every time shown.
In particular it is apparent that temperature we computed with the FEM-EV algorithm is far more 
diffusive than the temperature we computed with the other three algorithms.
Similarly the composition we computed with the FEM-EV algorithm is completely dissimilar from 
the composition field we computed with the other three algorithms.
In particular, the boundaries of the convection cells in both the upper and lower domains 
exhibit an unacceptably large amount of diffusion.

\subsubsection{Summary}
\label{Subsubsection:Summary}

In summary, we conclude that for computations in regimes with strongly stratified flow through 
the transition regime, namely $\mathrm{B} > \mathrm{B}_{cr}$ the VOF algorithm, or any other 
high quality interface tracking algorithm, is the most appropriate method for modeling the 
interface between the chemical compositions.
Our statement that the VOF algorithm will be superior to other algorithms in the transition 
regime is based on the first author's  experience with the VOF interface tracking algorithm in other
settings \cite{GHM-EGP:1994,GHM-EGP:1996,EGP-ASA-JBB-DLM-WJR:1997,LFH-PC-EGP:1991,LFH-EGP:2014}.
We are planning future work to explicitly demonstrate that this statement is true for this class of 
problems.

\subsection{Computations at $\mathrm{Ra} = 10^5$ with $ 0 \le \mathrm{B} \le 1.0$}
\label{Subsection:Computations at Ra eq 1e5 with 0 le B le 1}

In order to demonstrate the utility of this numerical methodology as a tool for studying 
thermochemical convection we computed a sequence of computations at $\mathrm{Ra} = 10^5$ with 
a sequence of buoyancy numbers $B \, = 0.0, \, 0.1, \, 0.2, \ldots, \, 1.0$.
This is covers the spectrum from the strongly stratified regime $B \ge  \approx \, 7.0$ through a 
transition regime $ \approx 0.6 \ge B \ge \approx 0.5$ to a regime that exhibits full kinematic mixing
$\approx 0.4 \ge B \ge \approx \, 0.3$, to the limiting case of 
uniform chemical composition $\Delta \rho = 0$ at $\mathrm{B}= 0.0$; i.e., steady single-layer 
thermal convection.

\begin{figure}[b!]
    \centering
    \includegraphics[width=\linewidth]{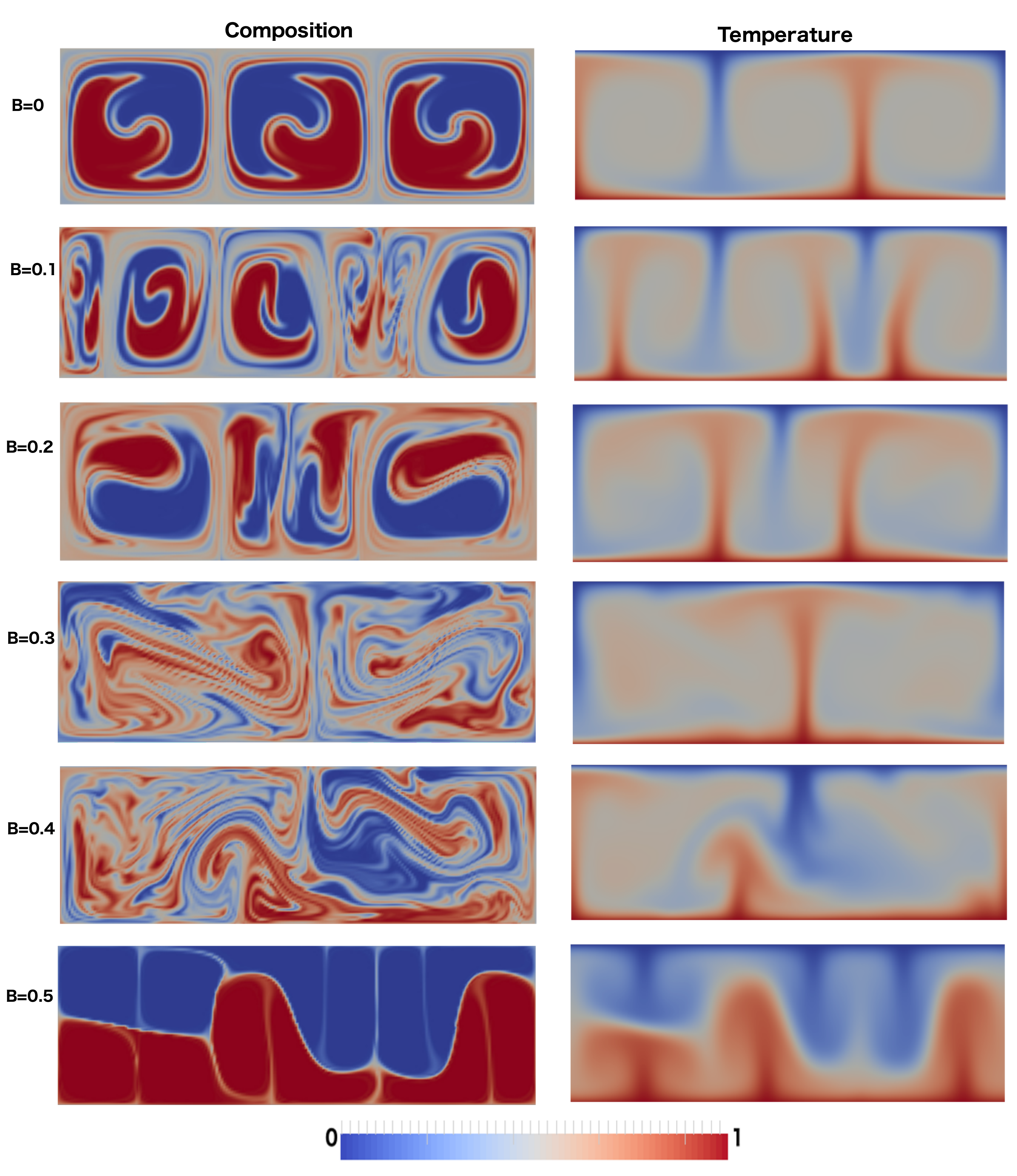}
    \caption{The composition and temperature at $t' = 0.15$ computed with the DGBP 
        advection algorithm at $\mathrm{Ra} = 10^5$ and
        $B = 0.0, \, 0.1, \, 0.2, \, 0.3, \, 0.4, \, 0.5$.
    }
    \label{Fig:BPDG_VARIABLE_BOUYANCY_COMPUTATIONS_1_to_5}
\end{figure}

\begin{figure}[t!]
    \centering
    \includegraphics[width=\linewidth]{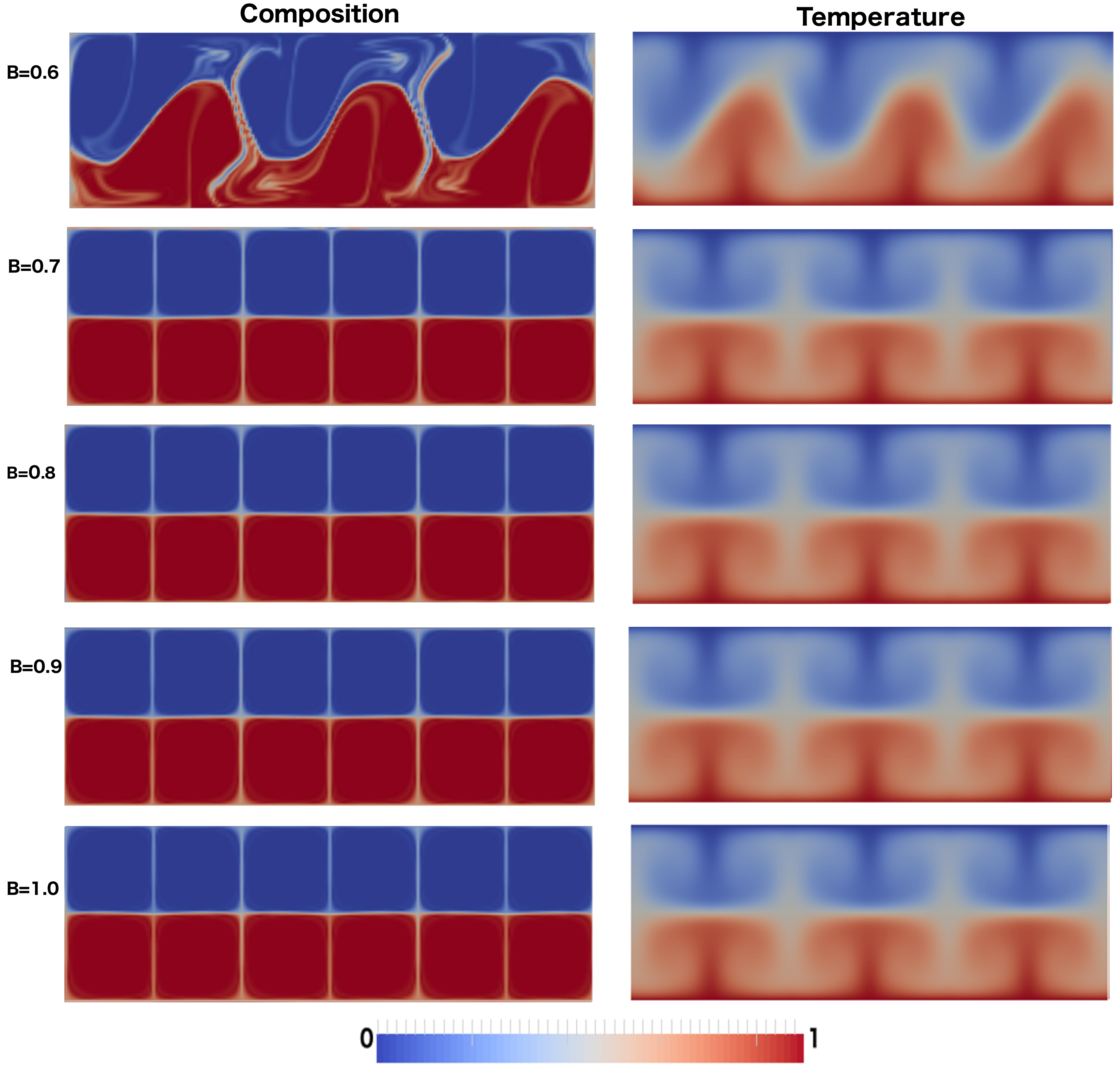}
    \caption{(A continuation of~Figure~\ref{Fig:BPDG_VARIABLE_BOUYANCY_COMPUTATIONS_1_to_5})
        The composition and temperature at $t' = 0.15$ computed with the DGBP composition 
        advection algorithm at $\mathrm{Ra} = 10^5$ and 
        $B = 0.6, \, 0.7, \, 0.8, \, 0.9, \, 1.0$.
    }
    \label{Fig:BPDG_VARIABLE_BOUYANCY_COMPUTATIONS_6_to_10}
\end{figure}

For this study we decided to use only one of the four algorithms presented above, as one 
might chose to do for a research study of thermochemical convection.
Our choice of algorithm was based on the following considerations.
\begin{enumerate}

  \item As demonstrated in    
      Figures~\ref{Fig:COMPOSITION_Combined_t_005}--\ref{Fig:TEMPERATURE_Combined_t_015}
      above, the FEM-EV algorithm produces computational results that are far too diffusive 
      to be regarded as reliable, especially as the other three methods show the same general
      computational results over a very long time period; i.e., approximately $1900$ time steps.

  \item For values of the Buoyancy number $\mathrm{B} < 0.5$, the compositional field undergoes 
        kinematic mixing on a scale for which it is inappropriate to use \textit{any} interface 
        tracing algorithm, including VOF.
        In particular we can rigorously quantify this by noting that the maximum curvature of the 
        interface $\sigma_{max}$ goes to infinity $\sigma_{max} \to \infty$, and hence from 
        the constraint
       we must also have the grid size $h  \to 0$.
        See \cite{EGP:2010a}, \cite{EGP:2010b} and \cite{EGP:2014} for details.
        
        \vskip 24pt

        The VOF algorithm is optimal for the range of values of $\mathrm{B}$
        for which the compositional density fields are stably stratified through values of regimes in which the two compositions do not overturn and undergo kinematic 
        mixing; i.e., $\mathrm{B} > \mathrm{B}_{cr}$ where $\mathrm{B}_{cr}$ is the value of 
        $\mathrm{B}$ at which the compositional density fields overturn and kinematic 
        mixing, begins to occur.
        As we have demonstrated in Figures~\ref{Fig:BPDG_VARIABLE_BOUYANCY_COMPUTATIONS_1_to_5} and~\ref{Fig:BPDG_VARIABLE_BOUYANCY_COMPUTATIONS_6_to_10}, when $\mathrm{Ra} = 10^5$
        we find $\mathrm{B}_{cr} > \approx 0.4$ 
        The DGBP and particle methods can be used in both the regimes, but they exhibit more numerical artifacts than the VOF algorithm for strongly stratified regime
        $B \ge  \approx \, 7.0$ and, more generally, in the transition or ``oscillatory regime''
        $\approx \, 7.0 \ge \mathrm{B} > \mathrm{B}_{cr}$.
        We plan future work to further substantiate this latter claim.      
        
   \item For $\mathrm{B} \, < \, \mathrm{B}_{cr}$ the DGBP and Particle algorithms are likely 
       to yield comparable computational results.
       For the following study we have chosen to use the DGBP algorithm.
       We have plans for future work that will include using the particle algorithm in ASPECT for 
       similar computations.
         
\end{enumerate}

Our numerical results at the final dimensionless time of $t'=0.15$ for
$\mathrm{B} = 0.0, \, 0.1, \, 0.2, \, \ldots , 0.5$  are displayed in 
Figure~\ref{Fig:BPDG_VARIABLE_BOUYANCY_COMPUTATIONS_1_to_5}, and for 
$B \, = \, 0.6, \, 0.7, \, \ldots, \, 1.0$ in
Figure~\ref{Fig:BPDG_VARIABLE_BOUYANCY_COMPUTATIONS_6_to_10}.
For $\mathrm{B} = 0.0$ the compositional boundary is completely passive since
$\Delta \rho = 0$.
Single layer thermal convection occurs with $\mathrm{Ra} = 10^5$.
At this Rayleigh number the convection is independent of time and passive kinematic mixing of 
the two layers occur in each cell.
Eventually the composition in the entire layer will approach $ C = 1/2$.
For $\mathrm{B} = 0.1$ and $0.2$ the compositional barrier impedes the single 
layer convection, but the passive kinematic mixing in individual cells continues to dominate.
For $\mathrm{B} = 0.3$ and $0.4$ the flow becomes much more chaotic but is dominantly single 
layer convection.
The chaotic behavior is attributed to a transition to a time dependent flow that enhances the 
kinematic mixing. 
For $\mathrm{B} = 0.5$ and $0.6$ the compositional boundary is distorted but the flow is 
basically two layer convection.
This range of $\mathrm{B}$ values give compositional structures that resemble the LLSVP 
structure of the Earth's mantle.
For $\mathrm{B} = 0.7$ and larger the compositional boundary blocks vertical flows and two 
layer thermal convection occurs.

\subsection{Relative Parallel Performance of the Algorithms}
\label{Section:Relative Parallel Performance of the Algorithms}

We conducted a strong scaling test to examine the degree to which the algorithms scale as the 
number of CPU processes or `cores' increases in a high performance parallel computing 
environment.
In a strong scaling test, a sequence of computations are made such that the size of the 
problem is fixed while the number of processor cores used for the computation increases 
\cite{MK-TH-WB:2012, MH-EH:2016, RG-EH-EGP-WB:2016}.

In our test we computed the problem described in 
Section~\ref{Subsection:Computations at Ra eq 1e5 with B eq 1 fixed}.
The grid resolution was fixed at $768 \times 256$ cells and we ran each method for $100$ time 
steps.
In the computation with particles there were initially 16 equally spaced particles per cell.
All of these computations were made on the high-performance computing cluster Maverick at the 
Texas Advance Computing Center. 

\begin{figure}[h!]
    \centering
    \includegraphics[width=0.7\linewidth]{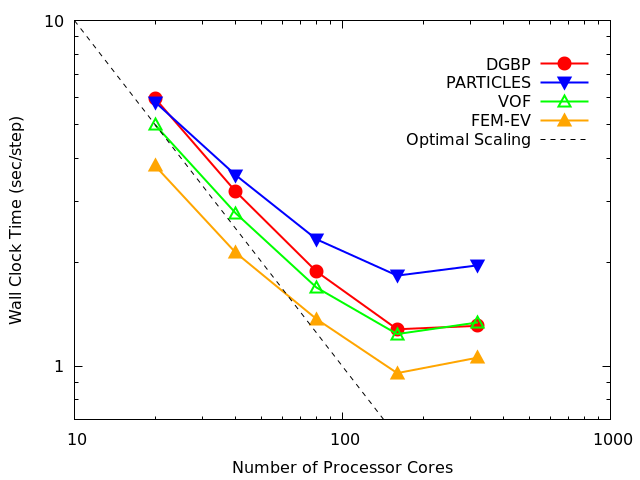}
    \caption{Strong scaling test for a fixed problem size of $768 \times 256$ equally spaced  
        square cells of size $h = 1 / 256$ with $\mathrm{B} = 1.0$ and $\mathrm{Ra} = 10^5$.
        The flattening of the curves as the number of cores exceeds 80 cores (= 4 nodes) is 
        expected. 
        This is an indication that the cost of communication between nodes is 
        becoming the dominant computational cost; i.e., there are too many nodes for the 
        problem size.}
    \label{Fig:STRONG_SCALING_WITH_POWER_LAW_FIT}
\end{figure}
\begin{figure}[h!]
    \centering
    \includegraphics[width=0.7\linewidth]{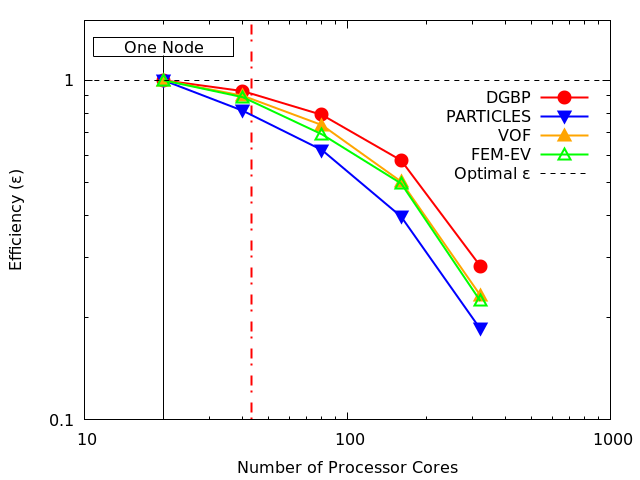}
    \caption{Strong scaling efficiency test for a fixed problem size.
        The 20 core case (1 node) is chosen to be the reference case in computing 
        efficiency.
        The dashed horizontal line represents optimal scaling ($\epsilon = 1$) and the dashed 
        red vertical line is the parallelization limit, after which the cost of communicating 
        between nodes begins to dominate the overall computational cost.
        The location of the vertical red line demonstrates that ASPECT, together any of the 
        four algorithms, is approximately 90\% efficient. 
            }
    \label{Fig:STRONG_SCALING_EFFICIENCY}
\end{figure}

The results are shown  in Figures ~\ref{Fig:STRONG_SCALING_WITH_POWER_LAW_FIT}
and~\ref{Fig:STRONG_SCALING_EFFICIENCY}.
For each computation, the elapsed wall clock time of the computation is measured and plotted. 
Figure~\ref{Fig:STRONG_SCALING_WITH_POWER_LAW_FIT} implies that of the different 
methods, the particle algorithm is more computationally expensive in comparison to the other 
algorithms. 
This is due to the additional cost associated with updating the location 
of the particles in time, and especially the communication cost incurred when passing 
particles across different MPI (Message Passing Interface) processes;
see also Figure~1 of~\cite{RG-EH-EGP-WB:2016} for a more detailed analysis of the parallel 
scalability of the particle algorithm in ASPECT.
Furthermore, Figure~\ref{Fig:STRONG_SCALING_WITH_POWER_LAW_FIT} shows that the DGBP and VOF 
algorithms have comparable wall clock times as we increase the number of processor cores.
Although, FEM-EV is by far the fastest method, given the failure of this algorithm to 
accurately model the underlying physics of the problem this is a false economy. 

We also examined the relative scalable efficiency of the four methods when we use them in 
ASPECT. 
The efficiency $\epsilon = 1$ is defined by 
\begin{align}
  \label{Def:STRONG_SCALING_EFFICIENCY}
   \epsilon = \frac{t_{ref}}{t}\frac{N_{ref}}{N_{Core}},
\end{align}
where $t$ is the elapsed time, $N_{Core}$ is the number of processor cores used, and 
$t_{ref}$ and $N_{ref}$ are the reference elapsed time and reference number of processor 
cores for each method \cite{MH-EH:2016}.
In this strong scaling efficiency test, we chose a single node consisting of 20 processor 
cores as the reference core case and the measured wall clock time for each method on a single 
node as the reference times.
The strong scaling efficiency results are shown in Figure~\ref{Fig:STRONG_SCALING_EFFICIENCY}.
Assuming that a local minimum problem size $10^5$ FEM Degrees of Freedom (DoF) per core is 
the threshold for this scalability, as was determined in~\cite{WB-CB-TH:2011, MK-TH-WB:2012}, 
then for our fixed problem size of $768 \times 256$ cells, we expect $43.3$ cores as the 
parallelization limit.
The strong scaling results shown in Figures~\ref{Fig:STRONG_SCALING_WITH_POWER_LAW_FIT} 
and~\ref{Fig:STRONG_SCALING_EFFICIENCY} demonstrate that each of the methods has good 
parallel scalability.
Similar strong scaling results for ASPECT have also been published in~\cite{MK-TH-WB:2012} 
and~\cite{RG-EH-EGP-WB:2016}. 

\section{Discussion}
\label{Section:Discussion}

\label{Subsection: A comparison of the FEM and DGBP advection algorithms}
Our numerical results in Section~\ref{Subsection:Computations at Ra eq 1e5 with B eq 1 fixed}
demonstrate the capabilities and limitations of the four algorithms we have used to model the 
motion of the compositional interface.
The numerical results at 
$\mathrm{Ra} = 10^5$ and $\mathrm{B} = 1$ show that the FEM-EV advection algorithm produces 
extremely diffusive temperature and composition fields.
Among these four algorithms, the FEM-EV and DGBP algorithms are both based on the 
Galerkin approach, which is a method for approximating the solution in the weak formulation. 
Since we applied the same time discretization for these two advection algorithms, the reason the 
results between the FEM-EV and DGBP advection algorithms  are so different must be due to the 
differences in spacial discretization.

The major difference between the standard FEM and the DG method is that the DG method 
allows for discontinuities between elements.
Therefore, it is well-known that DG is more suitable for problems with strong
discontinuities or large gradients \cite{CB-KG-SC:2000}, such as occurs at the boundary between 
the compositional fields in our problem.
In addition, in order to stabilize the numerical method, the FEM-EV algorithm uses an entropy 
viscosity technique, which is essentially adding an extra artificial diffusion term with 
variable artificial diffusivity $\nu_h^k (C)$ to  the original pure advection equation~\eqref{Eq:Nondimensional Vector Form of the Composition Equation} and resulting in
equations~\eqref{Eq:C_FEM} and~\eqref{Eq:The Entire FEM-EV Discretization}.

\begin{figure}[b!]
    \centering
    \includegraphics[width=\linewidth]{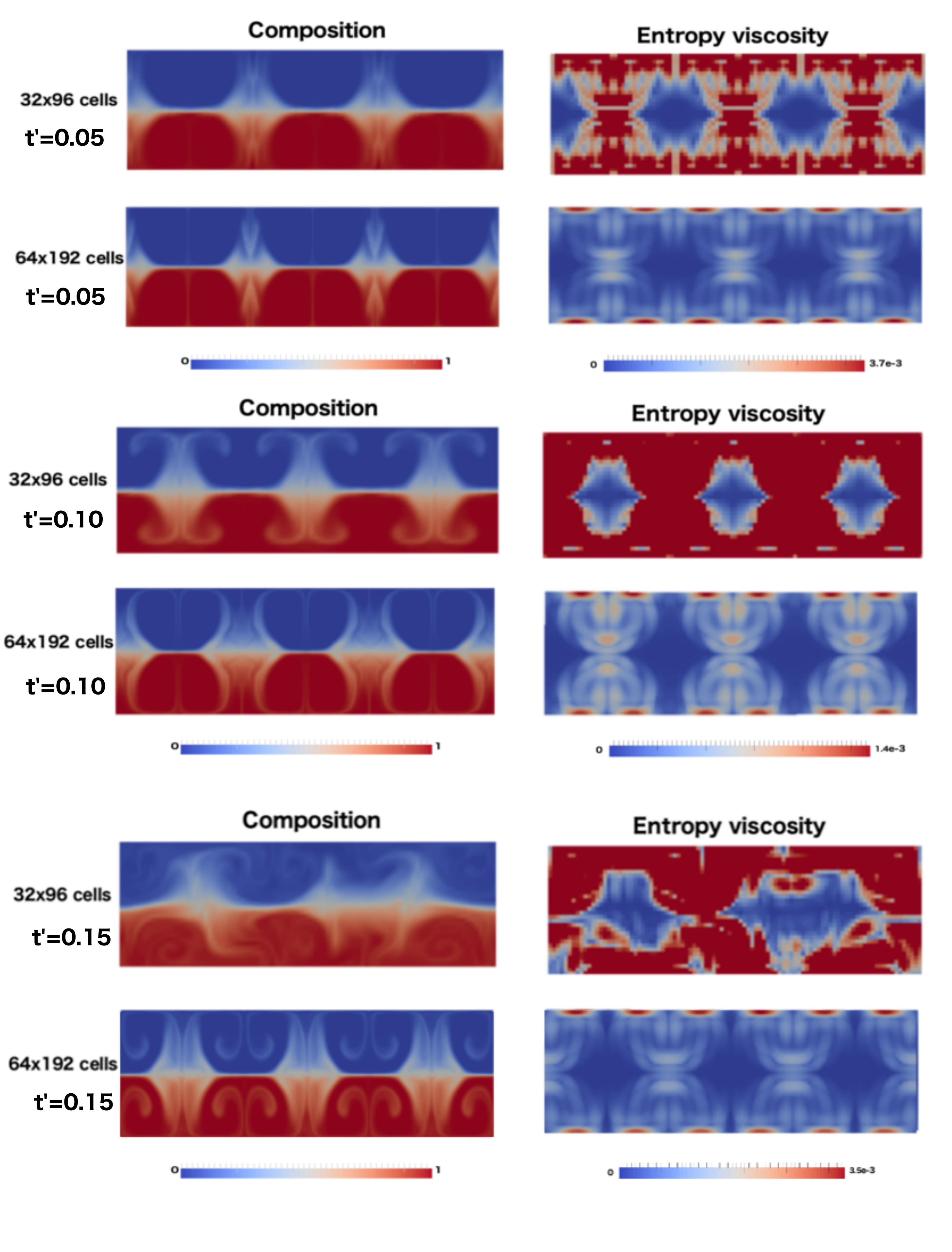}
    \caption{The composition and entropy-viscosity at time $t' = 0.05,\,0.10,\,0.15$ computed with 
             the FEM-EV composition advection algorithm for $\mathrm{Ra} = 10^5$ and
             $\mathrm{B} = 1$.
             }
    \label{Fig:Entropy_Viscosity_B_1}
\end{figure}

It is apparent from the definition of the entropy viscosity in~\eqref{Def:Entropy Viscosity} 
\cite{guermond2011entropy} that the value of entropy viscosity decreases as the grid size $h$ 
decreases.
Therefore, in order to reduce the amount of artificial diffusion, we have  to reduce the grid 
size $h$ for a fixed polynomial basis, say a $Q_2$ finite element. 
Figure~\ref{Fig:Entropy_Viscosity_B_1} shows that the maximum value of the entropy viscosity on  the grid is  greatly reduced from a computation on a grid of $96 \times 32$ cells to a computation 
on a grid of $192 \times 64$ cells for each time $t' = 0.05,\,0.10,\,0.15$ with the 
FEM-EV advection algorithm.
However, if the numerical computations need a very long time to compute, we cannot neglect the 
accumulated error due to the 
artificial diffusion term with a relative large grid size $h$. In other words, if the final 
computation time $t$ is fixed, we may can 
determine an appropriate small $h$ such that there is less accumulated artificial diffusion which 
will not significantly change the 
final solution. However, to have numerical solutions with the same accuracy, a much large grid size 
$h$ is already enough for the other three advection algorithms.
Therefore, the FEM-EV advection algorithm is the most diffusive method for a fixed $h$.

In contrast, in the DGBP advection algorithm we first discretize the problem by applying a standard 
DG method with an upwind monotone flux. 
This does not explicitly add an artificial diffusion term to the advection equation.
We then use a BP limiter in a post processing step in order to reduce or eliminate overshoot and 
undershoot in the DG solution near discontinuities in the composition variable $C$.
In \cite{XXZ-CWS:2010}, it is shown that the BP limiter does not reduce the accuracy of the 
original DG solution.
Also, numerical examples of the advection of non-diffusive fields in solid Earth geodynamics in 
\cite{YH-EGP-MIB:2016} show that the DGBP solutions preserve a much sharper boundary as compared 
to the FEM-EV method on the same mesh with the same grid size. 

However, for compared to the VOF interface tracking algorithm, $\mathrm{B} = 1$ the DGBP advection 
algorithm does not maintain a sharp compositional boundary, across which the two compositional 
fields do not mix. 
The ability to maintain sharp boundaries through time in a computation is likely to influence 
conclusions related to the entrainment of compositional signals from boundary layers into rising 
thermal plumes. 
The computational results in~\cite{YH-EGP-MIB:2016} indicate that coupling the DGBP method to 
Adaptive Mesh Refinement (AMR) will limit the computational cost as mixing proceeds, at least until 
the mixing leads to compositional heterogeneity at all scales.

\section{Conclusions}
\label{Section:Conclusions}

It is now widely accepted that compositional buoyancy plays an important role in mantle 
convection.
Heterogeneous regions of the mantle can be produced in a variety of ways.
One obvious source is the subducted lithosphere with the basaltic crust and a depleted 
complimentary mantle.
After subduction there are two possible end member results of the interaction of the layered 
lithosphere with the convecting mantle.
One is that the difference in densities between the subducted crust and the mantle results
in a segregated (layered) mantle.
The subducted crust could congregate in the deep mantle to create zero order layering.
The alternative is that the mantle is homogenized by kinematic mixing.

The basaltic crust and mantle do not mix by material diffusion except on scales of centimeters 
to meters on geological time scales.
Kinematic mixing is a tradeoff between the differential compositional buoyancy
$\Delta \rho \, g$ that tends to segregate the two components and the thermal convective 
buoyancy $\rho_0 \, \alpha \, \Delta T \, g$ that drives the kinematic mixing.
There are two controlling non-dimensional parameters, the Rayleigh number $\mathrm{Ra}$ that 
determines the vigor of thermal convection and the buoyancy ratio
$\mathrm{B} \, = \, \Delta \rho \, / \, \rho_0 \, \alpha \, \Delta T$ that determines the relative 
strength of the stabilizing chemical buoyancy $\Delta \rho \, g$ versus the driving thermal 
buoyancy $\rho_0 \, \alpha \, \Delta T \, g$.
It has  been the purpose of this paper to numerically study the segregation versus the kinematic 
mixing in a two component fluid for a range of buoyancy numbers $\mathrm{B}$ at a fixed Rayleigh 
number $\mathrm{Ra}$.
In order to do so, we first compared the accuracy and quality of the computational solution 
obtained from computing the compositional boundary with four alternative approaches in a strongly 
stably stratified regime  at a single, fixed, Rayleigh number; i.e., for extremely stable 
compositional segregation. 

Many previous computations of thermo-chemical convection have been carried out.
However, some numerical methods for modeling the compositional interface are too diffusive,
resulting in artificial mixing, which leads to inaccurate computational results. 
Since the compositional fields are non-diffusive, ideally the numerical methodology should be able  
to track the sharp boundary between each field for all values of $\mathrm{B}$ at a fixed Rayleigh 
number $\mathrm{Ra}$.
However, the extremely small length scales prevalent in fully developed kinematic mixing required 
us to choose a different numerical method than the method we found to be optimal for modeling the 
interface at values of $\mathrm{B}$ above that value $\mathrm{B}_{cr}$ at which the interface turns 
over and kinematics mixing develops.

In this article we describe in detail three alternative numerical methods to reduce numerical 
diffusion while modeling the advection of two compostional fields; namely, a Discontinuous Galerkin 
method with a Bound Preserving limiter (DGBP), a Volume-of-Fluid (VOF) interface tracking 
algorithm, and the advection of particles that carry a scalar quantity representing the location of 
each compositional field.
The first two of these methods are relatively new in the computational mantle convection 
community.
All three methods have been implemented in the open source Finite Element (FEM) code ASPECT for 
modeling processes that occur in mantle convection.
ASPECT is freely available from the Computational Infrastructure for Geodynamics.
We compare the performance of these three alternative methods with the advection method that was
first developed for modeling the advection of a compostional field in ASPECT.
This method is  based on a high-order accurate finite element advection algorithm with an 
artificial viscosity stabilization technique known as `Entropy Viscosity' (FEM-EV).
We use ASPECT to compute the velocity, pressure, and temperature fields associated with the 
underlying flow with each of these four methods. 

In order to examine the relative performance of these four methods, we have considered 
two-dimensional thermal convection in a fluid layer heated from below with an initial compositional 
barrier between the upper and lower half of the layer with Buoyancy number $\mathrm{B} = 1$ and 
Rayleigh number $\mathrm{Ra} = 10^5$.
This is a regime for which the initial stratification of the compositional fields persists 
indefinitely. 
The numerical results we computed with the FEM entropy viscosity-based method are far too 
diffusive to produce meaningful results.
We argue that this is, at least in part, due to the additional artificial diffusion that is added 
by the entropy viscosity method in order to stabilize the advection algorithm.
In contrast, the other three algorithms produce nearly identical temperature fields, even at 
relatively late times; e.g., after approximately 19000 time steps.

On the other hand, the compositional field computed with DGBP, VOF, and the particles have distinct 
differences. 
Computations made with the DGBP method exhibit some amount of each compositional field that is 
(numerically) entrained within the other compositional field and advects along the boundary of the 
convection cells in the wrong compostional domain.
The particle method exhibits a similar phenomenon, in which some particles representing the denser 
fluid are entrained in the upper, less dense fluid and advect along the boundary of the convection 
cells in the wrong compostional domain and, similarly, a small number of particles representing the 
less dense fluid are entrained in the lower, denser fluid and are advected along the boundary of 
the convection cells but in the wrong compostional domain.
Finally, we found that the VOF method maintains a sharp interface between the two compositions on a 
subgrid scale throughout the computation.
There is a small numerical error on the interface at two stagnation points that form at the 
intersection of four counter rotating convection cells. 
At these two points the flow is a classic shear flow centered on the interface, which is 
notoriously difficult to model. 
Our conclusion is that of the three algorithms we tested, in this regime the VOF interface tracking 
algorithm yields the most accurate numerical results.



We then computed the same problem in the same computational domain with Rayleigh number 
$\mathrm{Ra} = 10^5$ but for a range of buoyancy numbers $\mathrm{B} = 0.0,\,0.1,\ldots,\,1.0$ 
using the DGBP method in order to demonstrate the utility of this method when the stratified layer 
overturns and begins to mix.
This is a regime for which the VOF method is not suitable without an untenably large increase in the
resolution of the underlying grid.
In our judgment the DGBP algorithm was one of two approaches that we could have used to study the 
basic mixing versus segregation process; i.e., over a wide range of buoyancy numbers $\mathrm{B}$ 
for a given Rayleigh number.
We believe we could also have used the particle method for this study.
This will be the subject of future work. 

The results of our computations are summarized in 
Figures~\ref{Fig:BPDG_VARIABLE_BOUYANCY_COMPUTATIONS_1_to_5} 
and~\ref{Fig:BPDG_VARIABLE_BOUYANCY_COMPUTATIONS_6_to_10}.
There are basically four modes depending on the value of the buoyancy parameter $\mathrm{B}$.
Our results are for $\mathrm{Ra} \, = \, 10^5$ based on the full layer thickness.
With no chemical buoyancy $\mathrm{B} \, = \, 0$, a time independent cellular convection pattern 
develops and there is no mixing between adjacent cells.
For small chemical buoyancy, $\mathrm{B} \, = \, 0.1$ and $0.2$, this flow is maintained.
Kinematic mixing takes place in the individual cells but is relatively slow.
For $\mathrm{B} \, = \, 0.3$ and $0.4$ the flow is quite different.
The chemical buoyancy has resulted in unsteady flows that greatly enhance the kinematic mixing 
but do not block the vertical flows between the upper and lower layers.
At $\mathrm{B} \, = \, 0.5$ and $0.6$ there is sufficient chemical buoyancy to impede vertical flow 
between the two layers but the compositional buoyancy does not prevent a significant vertical
displacement of the compositional boundary.
Relatively little mixing between the layers is seen at this time and the structure resembles 
that of low shear velocity provinces.
For  $\mathrm{B} \, = \, 0.7$ to $1.0$ the chemical stabilizing buoyancy is sufficiently strong to 
block vertical flows between the upper and lower layers and the compositional boundary is essentially flat and thermal convection occurs in two 
independent layers.
\section*{Acknowledgements}
This work was supported by the National Science Foundation Grant Number 1440811.


\label{Section:References} 
\bibliography{ARXIV_DSF_PAPER_SINGLE}

\end{document}